\documentclass[preprint]{aastex61}
\usepackage{graphicx}
\usepackage{longtable}

\def\kms{$\textrm{km~s$^{-1}$}$}

\def\H2{H$_{2}$}

\def\roH2{$\rho_{\textrm{H}_2}$}
     
\def\MH2{M$_{\textrm{H}_2}$}

\submitjournal{AJ}

\shorttitle{Disks in isolated S0 galaxies} \shortauthors{Sil'chenko et al.}

\begin{document}

\title{The structure of stellar disks in isolated
lenticular galaxies.\footnotemark[1]\thanks{ Based on observations made with
the Las Cumbres Observatory telescope network.}}

\correspondingauthor{Olga Sil'chenko}
\email{olga@sai.msu.su,olgasil.astro@gmail.com}

\author[0000-0003-4946-794X]{Olga K. Sil'chenko} 
\affil{Sternberg Astronomical Institute, M.V. Lomonosov Moscow State University, Universitetsky pr., 13, Moscow, 119991 Russia}
\email{olga@sai.msu.su} 

\author{Alexei Yu. Kniazev}
\affil{South African Astronomical Observatory, PO Box 9, 7935 Observatory, Cape Town, South Africa}
\affil{Southern African Large Telescope Foundation, PO Box 9, 7935 Observatory, Cape Town, South Africa}
\affil{Sternberg Astronomical Institute, M.V. Lomonosov Moscow State University, Universitetsky pr., 13, Moscow, 119991 Russia}
\email{akniazev@saao.ac.za}

\author{Ekaterina M. Chudakova} 
\affil{Sternberg Astronomical Institute, M.V. Lomonosov Moscow State University, Universitetsky pr., 13, Moscow, 119991 Russia}
\email{artenik@gmail.com} 

\begin{abstract}
We have obtained imaging data in two photometric bands, $g$ and $r$, for a sample of 42 isolated lenticular galaxies with the Las Cumbres
Observatory one-meter telescope network. We have analyzed the structure of their large-scale stellar disks.
The parameters of surface brightness distributions have been determined including the radial profile shapes and disk thicknesses.
After inspecting the radial brightness profiles, all the galaxies have been classified into pure exponential (Type I),
truncated (Type II), and antitruncated (Type III) disks. By comparing the derived statistics of the radial profiles shapes with our
previous sample of cluster S0s, we noted a prominent difference between stellar disks of S0s galaxies in quite rarefied
environments and in clusters: it is only in sparse environments that Type II disks, with profile truncations, can be found.
This finding implies probable different dynamical history of S0 galaxies in different environments.
\end{abstract}


\section{Introduction}  

The structure of a lenticular galaxy implies combination of a bulge and of a large-scale stellar disk,
hence being quite similar to that of spiral galaxies. However, stellar disks of S0s are usually red and
smooth and do not contain clumpy current star formation. Due to this (dis-)similarity lenticular galaxies
are commonly thought to be descendants of spirals devoid of gas. It is easy to remove the gas from the galactic
disk when a galaxy enters dense environments; many mechanisms for this purpose are proposed: ram pressure
\citep*{gunngott,quilis} or static pressure \citep{cowieson} by dense hot intergalactic medium,
tidal effects -- from the whole cluster \citep{byrdvalt}, pair collisions \citep{spitzerbaade},
harassment \citep{moore96}, slow fly-by \citep{icke,bekkicouch}, starvation \citep*{larsons0} etc.

But lenticular galaxies inhabit not only clusters, though they are the most numerous population there. A lot of S0s belong
to loose groups, and some lenticulars are even quite isolated. It seems natural to suggest that they may have different
evolutionary paths. Then the structure of their disks may reflect different dynamical evolution of S0s in different
environments.   The term `structure' refers here to both radial and vertical surface-brightness distributions.

Concerning the radial structure of galactic stellar disks, it is presently established that the slope of their radial 
stellar surface-brightness decrease can be typically described by a piecewise exponential function. Now three main types
of the radial brightness profiles are recognized: they can be fitted by a single-scale exponential over the whole extension 
of a stellar disk \citep{freeman}, or by an exponential law with truncation at some radius as firstly mentioned in \citet{freeman} 
and later classified as 'breaks' by \citet{vdkruit_searle}, or with two exponential segments, the outer exponential law
having a larger scalelength -- so called antitruncated disks \citep*{n5533,n80,erwin05}. 
After the SDSS statistics analysis by \citet{pohlen_trujillo}, these three types of surface-brightness profiles 
become to be numbered as follows: single-scale exponential disks are Type I, truncated disks are Type II, and antitruncated 
disks are Type III. The hot topic of discussion is if the shape of a stellar surface-brightness (density) profile is
an initial condition of the galactic disk formation, or there exists some dynamical transitions between
them. Firstly \citet*{erwin12}, and later \citet{pranger} and \citet*{lcogt_clust}, have reported a hint on the environment 
effect concerning the profile-shape statistics for early-type disk galaxies in clusters: there is a deficit
of Type II profiles compensating by an excess of Type I profiles in these dense environments. Trying to give an evolutionary basis to
this discovery, \citet{clarke_db} proposed a dynamical mechanism to transform truncated disks into a single-scale exponential
one during the infall of a galaxy into cluster environment: gas stripping by ram pressure and enhanced
stellar radial migration together provide respective changes in a stellar disk structure. However, this
mechanism works only for {\it gas-rich late-type} spirals and requires subsequent transformation of spirals into
lenticulars.  Concerning the origin of the most common, Type III disks of S0s, there were dynamical simulations 
proposing transformation of a single-scale exponential disk into an antitruncated one  \citep[e.g.][]{younger,borlaff14}.
Comprehensive cosmological simulations reveal the most violent dynamical evolution exactly for the Type-III disks:
in the frame of LCDM-models they are formed by strong radial migration as well as by concentration of freshly accreted stars
in their outermost parts \citep{ruizlara}.

Apart from the radial profile type, in this paper we consider another structure characteristics of galactic disks -- their thickness.
It is important to get estimates of the disk thicknesses for different disk radial-profile types because it will allow us to limit
possible dynamical mechanisms shaping the large-scale stellar disk structure of real lenticular galaxies.
In particular, dry minor mergers which have been elected by \citet{younger} to form an antitruncated
surface-density profile, are also expected to thicken stellar disks by increasing their vertical stellar
velocity dispersion \citep*{walker96}. Observational estimates of galactic disk thicknesses are rare:
individual estimates of stellar disk thickness were made directly only for galaxies seen 
edge-on \citep*[e.g.][]{mosenkov10,mosenkov15,comeron_edgeons4g}. However it is rather indirect
to discuss radial and/or azimuthal structure of the galaxies seen strictly edge-on.
We \citep{thickmeth} have invented a novel method allowing to estimate an individual thickness
of an exponential (or piecewise exponential) stellar disk seen under arbitrary
inclination; only strictly edge-on or strictly face-on disks cannot be analyzed by our method.
With our method, we have already made some efforts, with the aim to compare disk thicknesses among the samples
of various types of radial surface-brightness profiles for early-type disk galaxies in groups
\citep*{thickmeth,lcogt1} and in clusters \citep{lcogt_clust}. In the present paper we continue 
to apply our method of estimating thicknesses in individual galactic disks to a sample of isolated S0 galaxies.
In Section 2 we describe the sample, in Section 3 we give details of our observations and of our approach to
the stellar disk structure characterization, in Section 4 we present our quantitative results and discuss them,
and in Section 5 we conclude.

\section{Sample}

\begin{figure}
\centering
\includegraphics[width=0.45\textwidth]{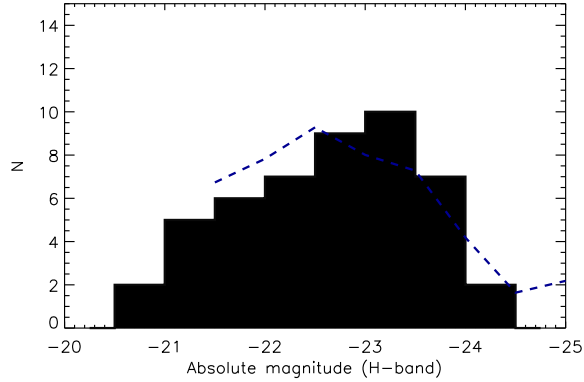}
\caption{The distribution of the galaxy absolute magnitudes in the $H$-band for our sample;
the $M_H$ are taken from NED. The dashed line overposed represents the luminosity function of
the volume-limited sample of early-type galaxies from \citet{atlas3d_1}.}
\label{mh_distrib}
\end{figure}

The aim of our study is to inspect photometric structure of isolated lenticular galaxies. 
The criterium of isolation, $II$, has been defined by Karachentsev and Makarov with coauthors
\citep{makkar,karlog}; it is based on the consideration of mutual gravitational effect of all possible
pairs of galaxies. The parent sample of rather isolated lenticular galaxies 
compiled by \citet{katkov_disser} included 281 S0 galaxies within the volume of $v_r<$4000~\kms\ selected through 
the criterium of isolation index $II>2.5$. This choice of $II$ means that the mass of every sample galaxy has to be increased
by a factor of 2.5 -- or the nearest brighter galaxy mass has to be increased by a factor of 2.5 -- to join 
the sample galaxy into a gravitationally bound ensemble with another galaxies. The consideration was limited by
galaxies having the K-band magnitude fainter by 2.5 mag than our target galaxies. In other words, presently our
galaxies are gravitationally unbound to any brighter galaxy or to any fainter galaxy within 10\%\ of its mass.
In the frame of our photometric project we have obtained imaging data 
for 42 southern-sky S0 galaxies from this parent sample; in order to be able to apply our method of disk thickness 
estimation we avoided strictly edge-on disks. The galaxies included into the sample are classified as S0 or S0/a 
in the NASA/IPAC Extragalactic Database (NED) or in the HyperLEDA. One galaxy, PGC~11756, is classified in HyperLEDA
as E; but it is an obvious mistake since we have found that it lacks bulge (spheroid) at all. Additional check of the presence
of a large-scale stellar disk in the galaxies selected for our analysis was made after deriving
the surface brightness profiles by the method described in the next Section; we suggest that the main difference
between S0 and an elliptical is the presence of a smooth large-scale disk. For every galaxy we started with a fit
of the outer parts of the surface-brightness profiles by exponentials and proved that there exists at least
one segment of a profile which lacks any systematic deviations from the exponential law over a radial range of two
exponential scalelengths. We took this criterium -- namely, an exponential law validity over a two-scalelength radial range --
from \citet{freeman}: it was formulated in this classical work as a necessary property of exponential stellar disks.
The range of luminosities of the isolated S0s in our study is found to extend from $M_H=-21$ to $M_H=-24.3$.
We have compared the distribution of the NIR absolute magnitudes of our galaxies (Fig.~\ref{mh_distrib}) 
with the luminosity function of the volume-limited sample of nearby early-type galaxies from \citet{atlas3d_1} 
and have assured that they agree well: the Kolmogorov-Smirnov (K-S) test has shown the difference statistics of
$d_{max}=0.072$, or $\lambda =0.42$, that means that the distributions are the same with the probability greater than 99\%.
With our sample of isolated lenticulars we probe all the luminosity range of non-dwarf S0 galaxies,
so we may hope that our sample of isolated S0s is representative.

We want to compare the results obtained here for the sample of isolated lenticular galaxies to those for
the sample of 60 cluster lenticular galaxies which have been studied by us earlier \citep{lcogt_clust}.
The cluster S0s, in the same range of luminosities, were taken by us in 8 southern clusters of galaxies,
spanning a range of masses (X-ray luminosities) but all being not too far from us -- within $D=71$~Mpc.

\section{Observations and data analysis}

The photometric observations of the sample of isolated S0s have been undertaken with the 
Las Cumbres Observatory (LCO) robotic telescope network \citep{lcogt} during 2016--2018.
The LCO network includes two 2-m telescopes, ten 1-m telescopes, and
ten 40-cm telescopes -- in total 22 telescopes distributed among 7 different observatories.
All our observations were done with the LCO one-meter telescopes equipped by standard Sinistro cameras
intended to acquire image frames. Such camera contains a 4000$\times$4000 CCD. With the
physical pixel size of  15$\mu$m and standard 1$\times$1 binning, we get a scale of 0.389 arcsec
per pixel, and every single frame covers an area of 26$\times$26 arcmin.
Each Sinistro camera is equipped by 21 different filters,
of which we used the $g$ and $r$ broad-band filters of the Sloan survey photometric system.
The log of the observations is given in the Appendix.

During the observations, there were no dedicated exposures of photometric standards, so we calibrated our
images by three different ways:
\begin{itemize}
\item{by using the HyperLEDA\footnote{\textrm{http://leda.univ-lyon1.fr}} 
aperture photometry data -- the Johnson-Cousins $B(V)R$ aperture measurements for every galaxy,
mostly based on the compilations of the photometric survey of the southern sky \citep{esolv} which  
we transformed into the $gr$-system with the interrelations found by \citet*{sdsscal};}
\item{for the galaxies covered by the SDSS survey (NGC~270, NGC~1656, NGC~6014, UGC~5745) we have
used the $gr$-photometry of nearby stars from the SDSS/DR9 public data archive \citep{sdss_dr9} as standards;} 
\item{also we have used $gr$-photometry of the neighboring stars from the Pan-STARRS1/DR1 \citep{panstarrs_1,panstarrs_2}
public data archive as standards for the galaxies with declinations larger than --30\degr.}
\end{itemize}

The images were initially reduced by the LCO pipeline which performs bias subtraction and flat-fielding
of individual frames. Then we co-added the images made in the same filters for every galaxy, proceeding
cosmic hit cleaning simultaneously, and estimated the sky background distributions in the aggregate frames.
The sky background level for every galaxy was estimated over several large, by $51 \times 51$ pixels, empty square areas,
taken in at least eight different directions from the galaxy. Then the sky background estimates were averaged,
or, in the case of detected gradient over the frame, interpolated linearly onto the galaxy position.

\begin{figure*}[bpt!]
\centering
\begin{tabular}{c c}
 \includegraphics[width=8cm]{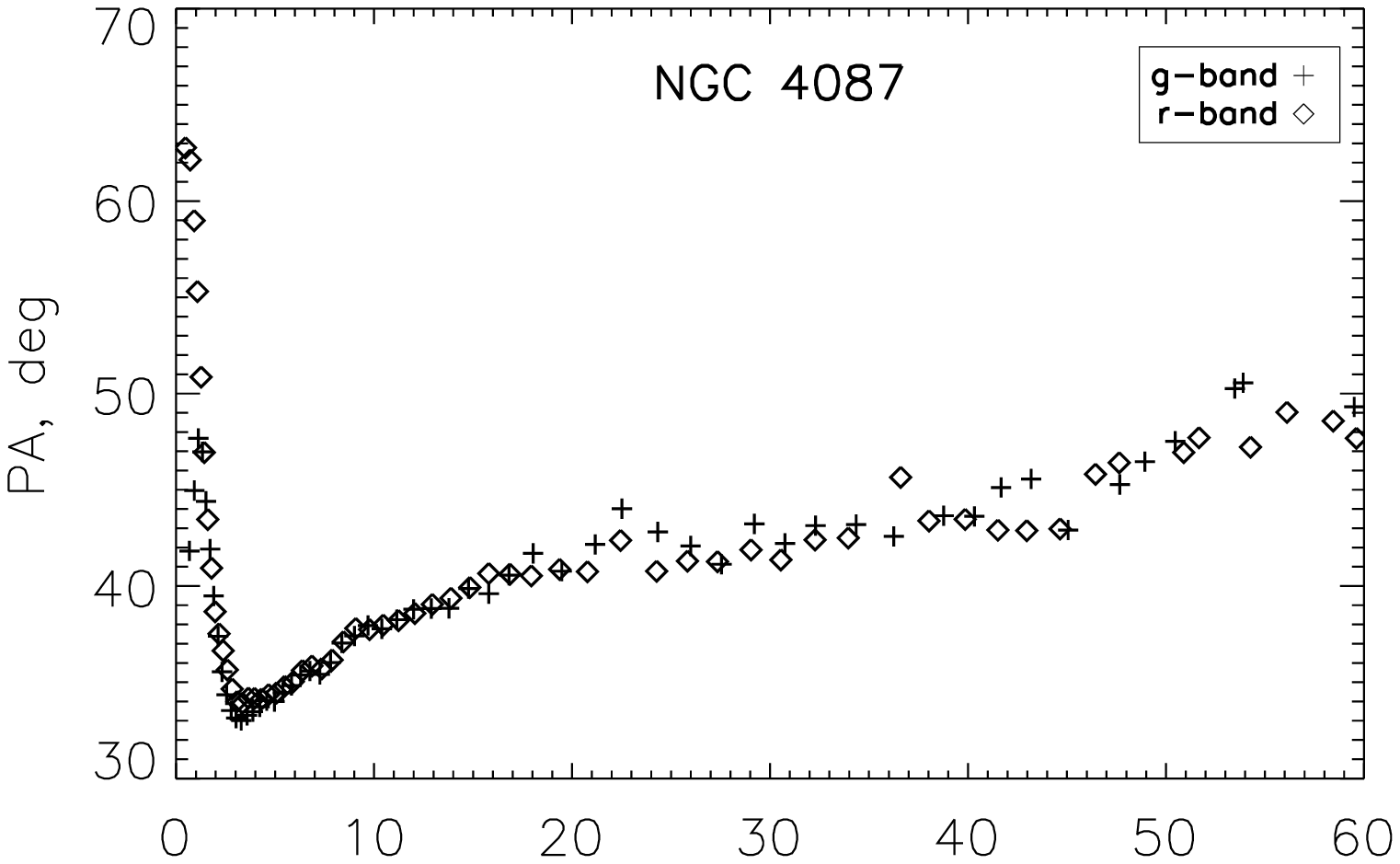} &
 \includegraphics[width=8cm]{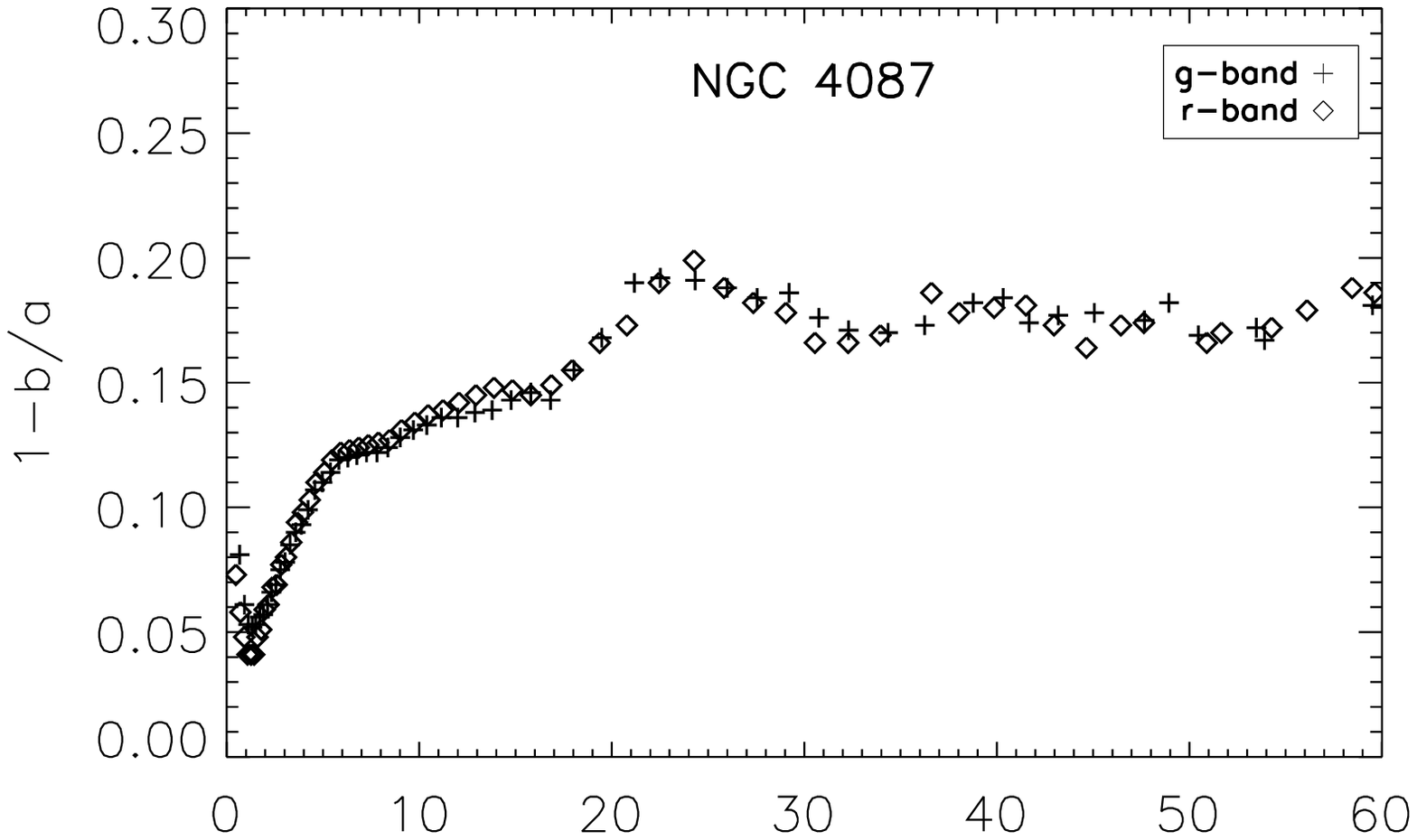} \\
 \end{tabular}
 \includegraphics[width=10cm]{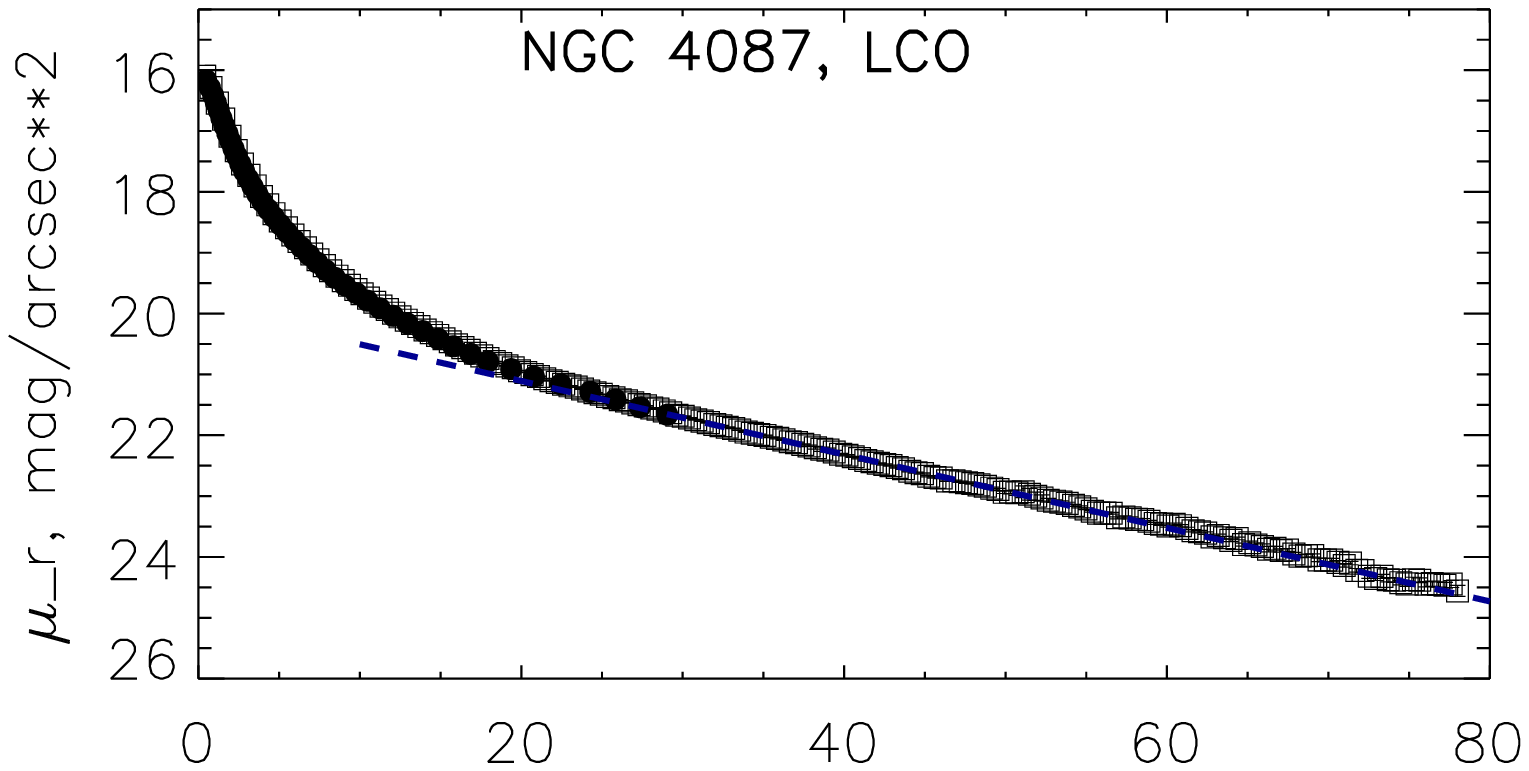} \\
 \vspace{0.3cm}
\caption{An example of a single-scale exponential surface-brightness profile (Type I):
NGC~4087. The upper row shows the results of the isophote analysis in two bands, the bottom plot is an
azimuthally-averaged surface-brightness profile in the $r$-band. In the surface-brightness profile the
black dots present the measurements azimuthally averaged with running ellipticity and
orientation of the elliptical aperture, while open squares with error bars -- those with fixed
orientation of the elliptical aperture corresponding to the line-of-nodes position angle
and inclination of the disk (see the text).}
\label{type1}
\end{figure*}

\begin{figure*}[bpt!]
\centering
\begin{tabular}{c c}
 \includegraphics[width=8cm]{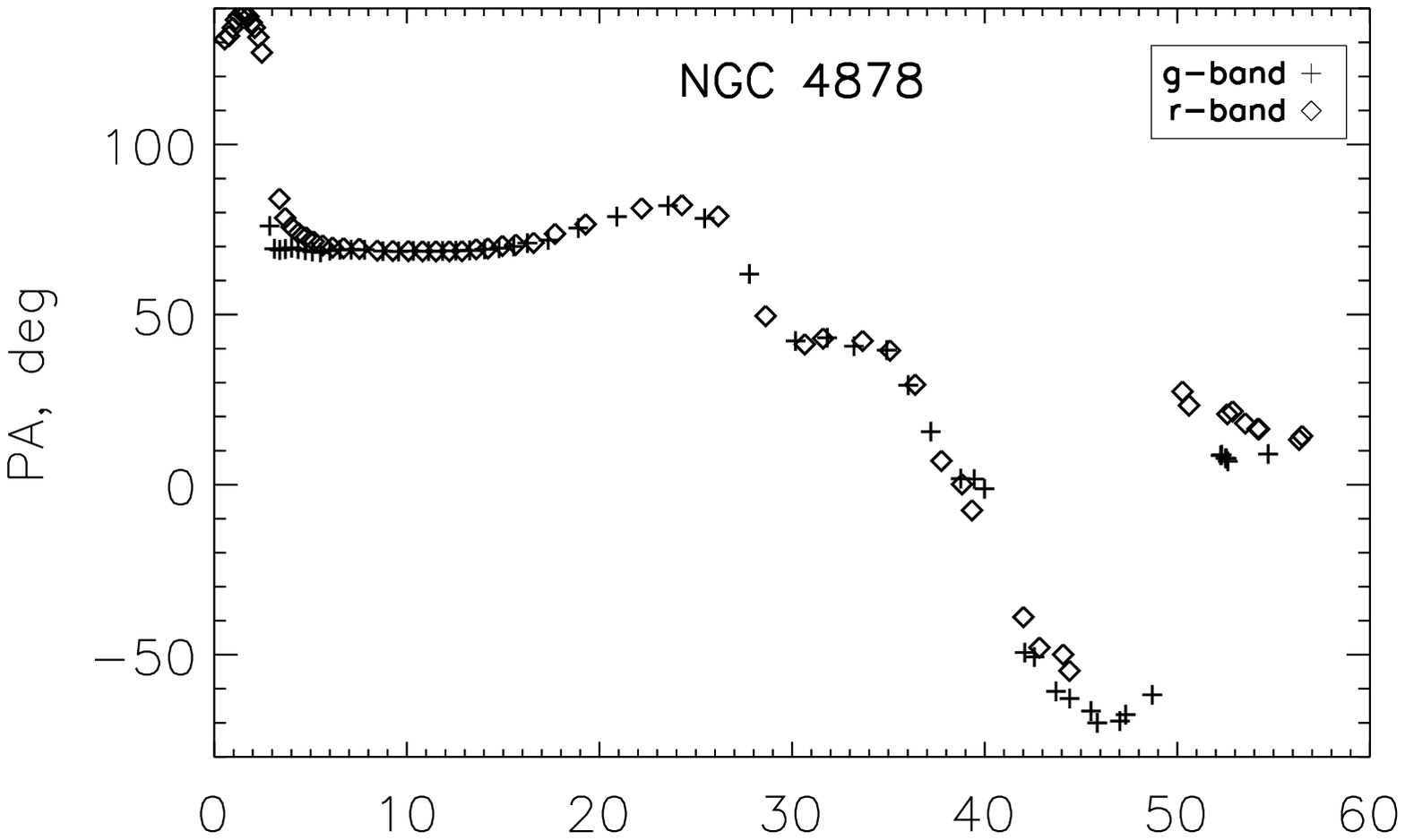} &
 \includegraphics[width=8cm]{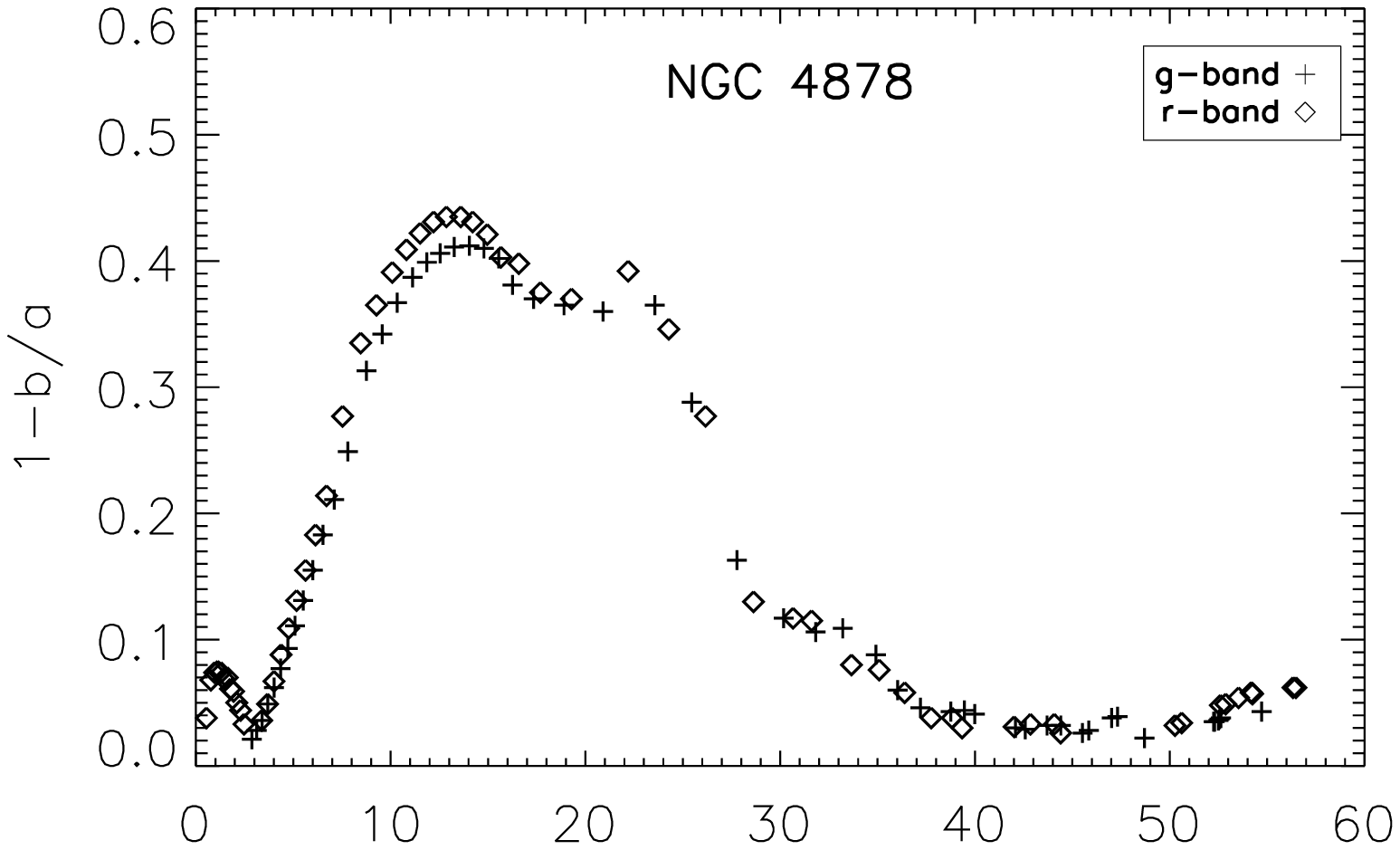} \\
 \end{tabular}
 \includegraphics[width=10cm]{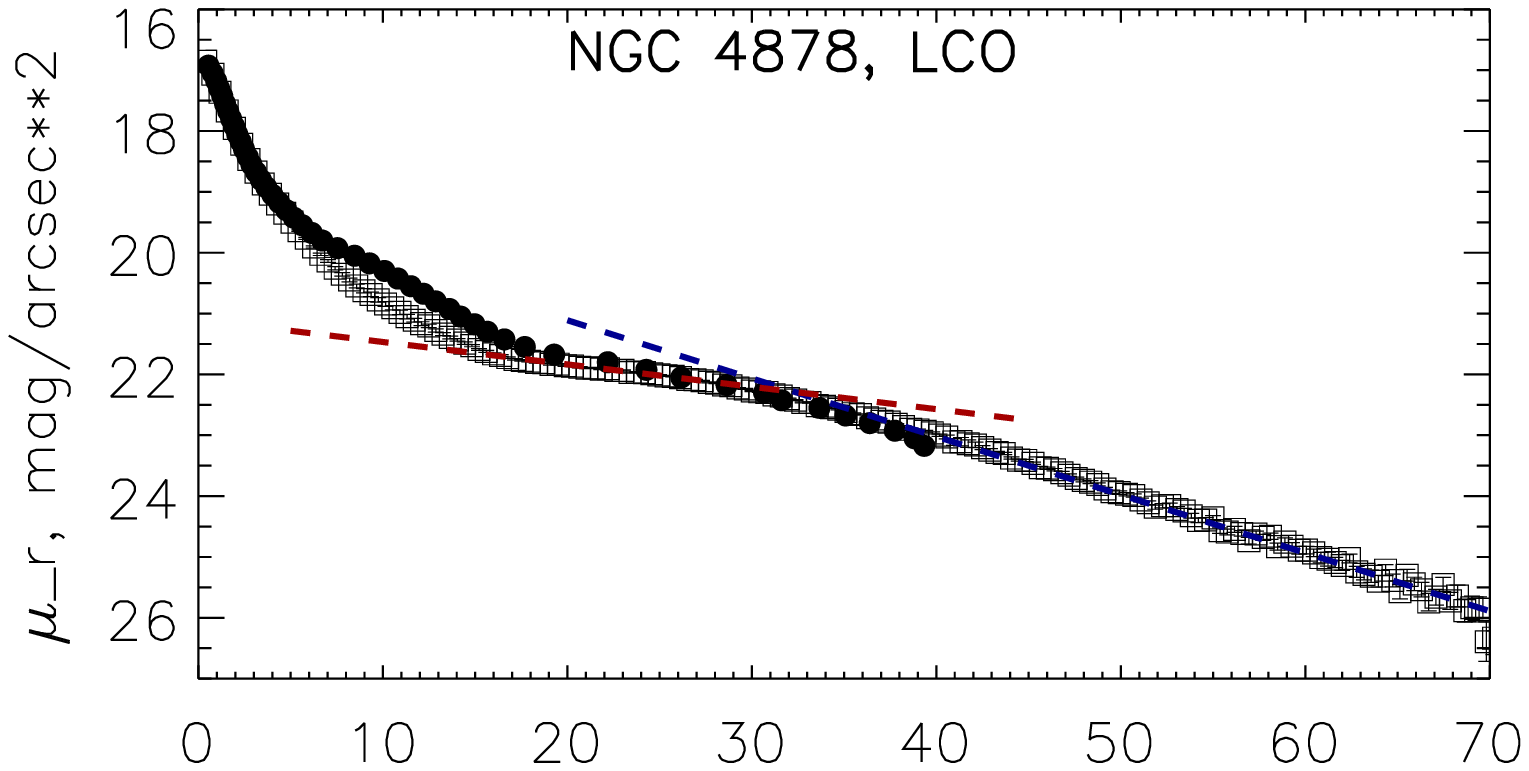} \\
 \vspace{0.3cm}
\caption{An example of a truncated piecewise exponential surface-brightness profile (Type II):
NGC 4878. The upper row shows the results of the isophote analysis in two bands, the bottom plot is an
azimuthally-averaged surface-brightness profile in the $r$-band. In the surface-brightness profile the
black dots present the measurements azimuthally averaged with running ellipticity and
orientation of the elliptical aperture, while open squares with error bars -- those with fixed
orientation of the elliptical aperture corresponding to the line-of-nodes position angle
and inclination of the disk (see the text).}
\label{type2}
\end{figure*}

\begin{figure*}[bpt!]
\centering
\begin{tabular}{c c}
 \includegraphics[width=8cm]{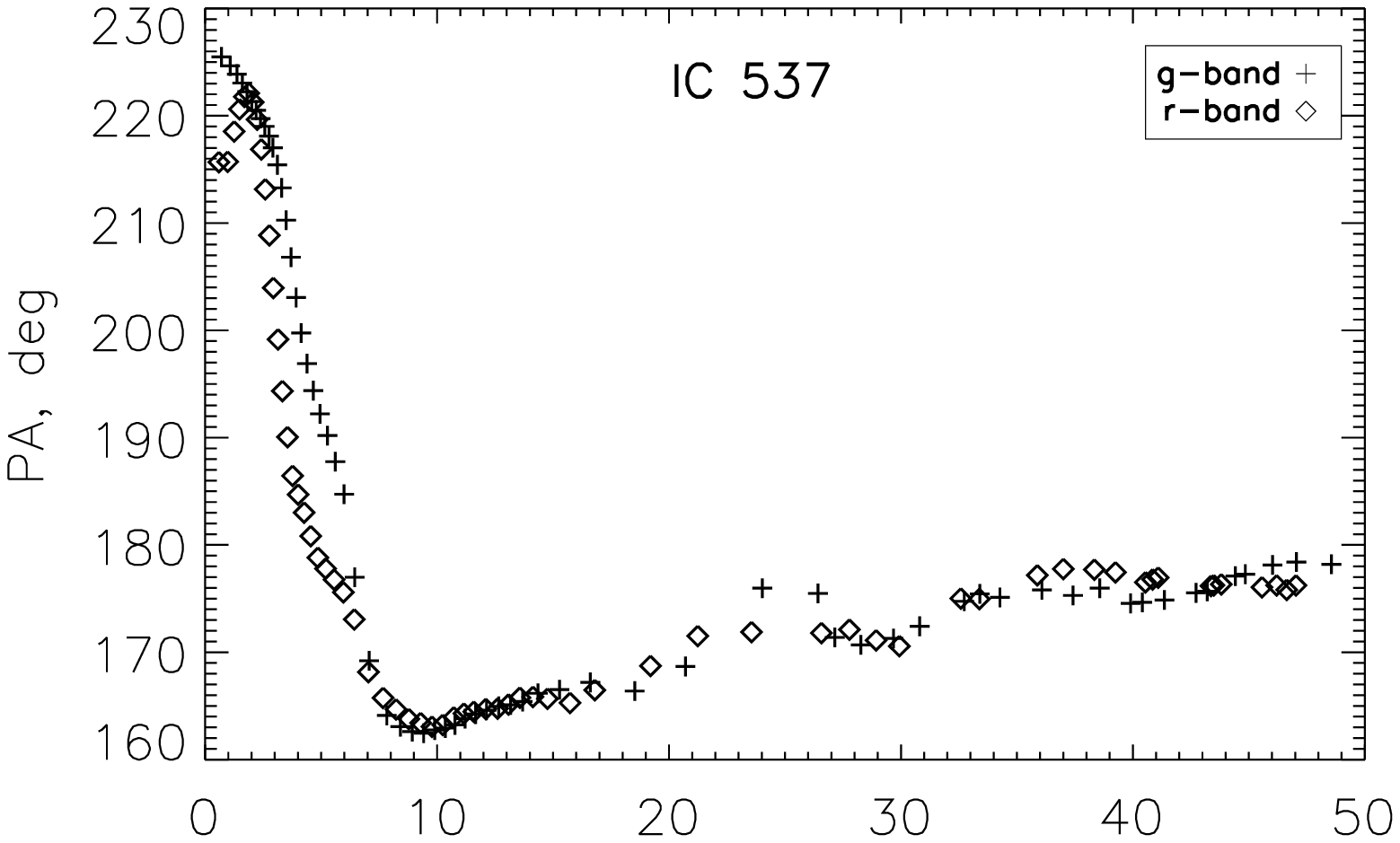} &
 \includegraphics[width=8cm]{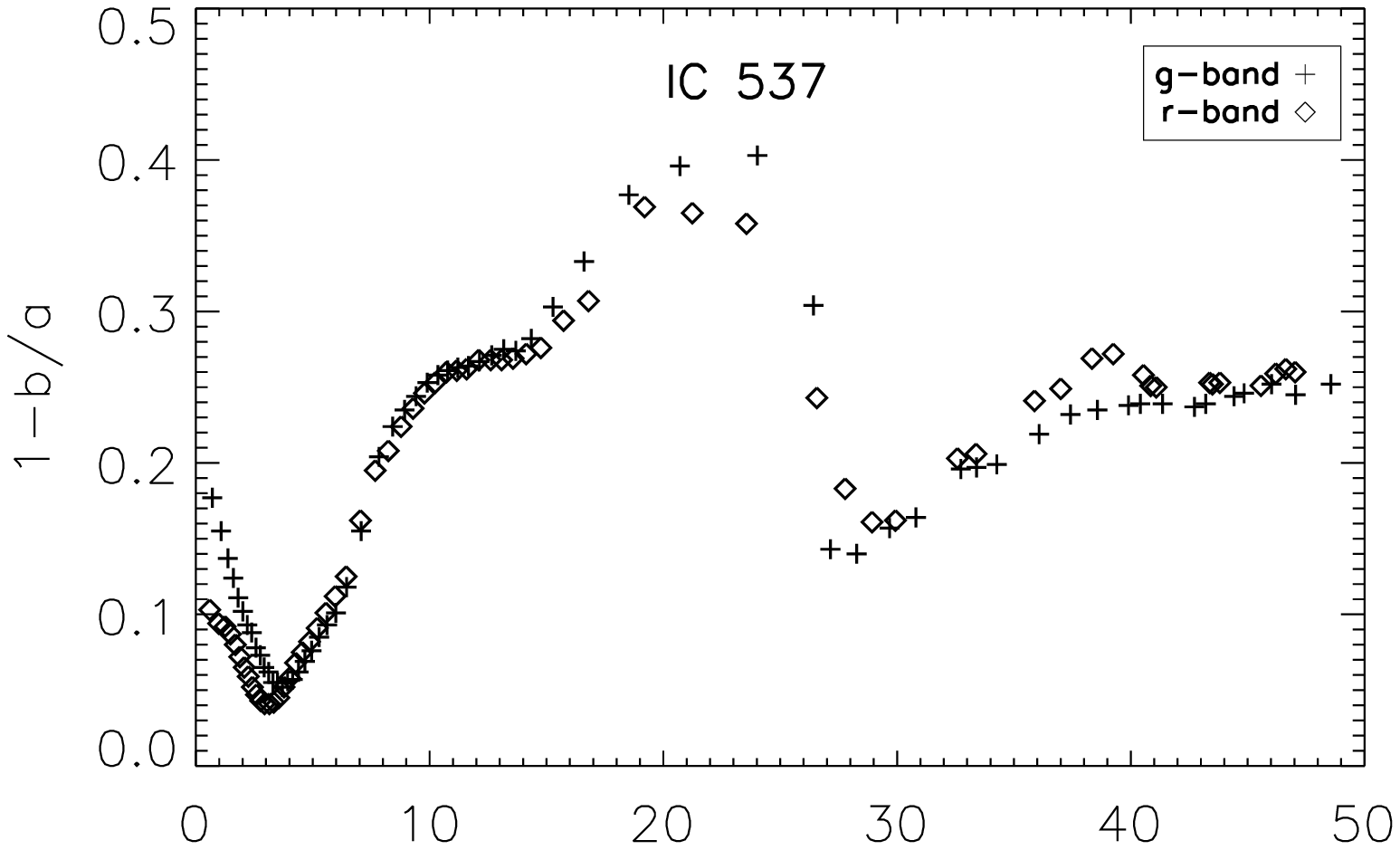} \\
 \end{tabular}
 \includegraphics[width=10cm]{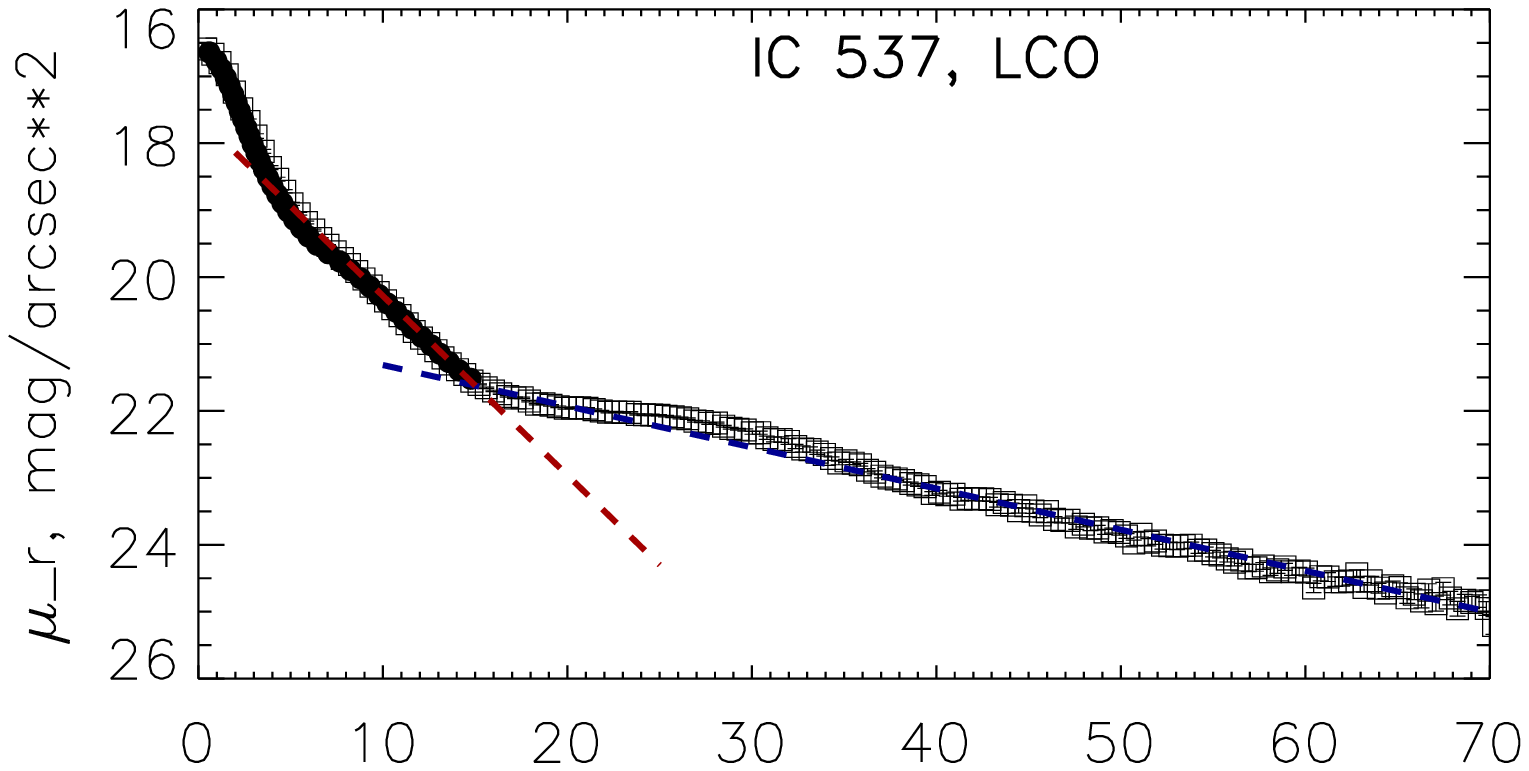} \\
 \vspace{0.3cm}
\caption{An example of a two-tiered (antitruncated) exponential surface-brightness profile (Type III):
IC 537. The upper row shows the results of the isophote analysis in two bands, the bottom plot is an
azimuthally-averaged surface-brightness profile in the $r$-band. In the surface-brightness profile the
black dots present the measurements azimuthally averaged with running ellipticity and
orientation of the elliptical aperture, while open squares with error bars -- those with fixed
orientation of the elliptical aperture corresponding to the line-of-nodes position angle
and inclination of the disk (see the text). The outer segment of the disk contains a ring at the radius
of 30\arcsec .}
\label{type3}
\end{figure*}

After obtaining the flat-fielded and sky-subtracted images, we made then isophotal analysis for every
galaxy by using the analog of the ELLIPSE/IRAF algorithm and derived radial profiles of the isophote
ellipticities and major-axis position angle by going from the galaxy center outward with the logarithmic brightness
step of 0.05. By assuming that the large-scale stellar disks of our S0s were flat and did not suffer warps, for every galaxy
we fixed a radius where the isophote ellipticity stopped to rise steeply and reached a plateau. We
suggest that the flat disk dominates in the total surface brightness of an outer region of a galaxy
beyond this radius. To derive azimuthally averaged surface-brightness profiles of the disks,
we selected the parameters of the elliptical apertures intending to characterize a round disk projection onto the sky plane, by exploring
the isophote ellipticity and major-axis position angle found just at this radius marking the rise of the disk-dominated area.
Then we averaged the surface brightnesses over the elliptical rings by going outward with some steps along the radius.
Finally the azimuthally averaged surface-brightness profiles of the stellar disks were obtained.
For the innermost regions, where the ellipticity and major-axis position angle of the isophotes changed strongly along the radius,
we averaged the surface brightness in the elliptical rings with the orientation and
shape modifying radially in accordance with the results of the isophote analysis. After deriving azimuthally averaged
surface-brightness profiles, we started to fit them by an exponential function selecting initially the radial range between
the outermost point exceeding the sky level by an rms sky-level scatter value and the innermost radius
where the profile points still obey to the exponential law fitted. The quality of the fit was taken to be good
if the rms scatter of the points around the fitting line was found to be within typical errors of the individual measurements.
If we found an inner radius where the azimuthally averaged surface-brightness profile started to deviate systematically
up or down from the fitted exponential law by a value exceeding the rms scatter of the observational points around
the fitted exponential law and if this radius was still within the disk-dominated radial range characterizing by the stable high
isophote ellipticity, we concluded that the profile is not of the Type I, and fitted another exponential segment into the inner part
of the azimuthally averaged surface-brightness profile. By appying this procedure to the whole sample, we have divided the total
galaxy list into three subsamples: the S0s with Type-I profiles, the S0s with Type-II profiles, and the S0s with Type-III profiles.
Figures~\ref{type1}, \ref{type2}, and \ref{type3} demonstrate some examples of all three types of the disk 
surface-brightness profiles in our sample.

{\small
\startlongtable
\begin{deluxetable}{lclrcrr}[t]
\scriptsize
\tablecolumns{7}
\tablewidth{0pc}
\tablecaption{The galaxies studied photometrically with the LCO network.\label{listgal}}
\tablehead{
\colhead{Galaxy} & \colhead{Type\tablenotemark{a}}  & \colhead{$M_H$\tablenotemark{a}} &
\colhead{$R_{25}$,$^{\prime \prime}$\tablenotemark{b}} & \colhead{Profile type} & \colhead{Bar/ring?}  & \colhead{Color features} }
\startdata
ESO 003-001 & (R$_1$)SB(rs)0/a & --23.74 & 46.5 & III & bar, ring & blue knot in 5'' to NW \\
ESO 040-002 & SA0$^-$: & --23.98 & 39 & III & \nodata & \nodata \\
ESO 052-014 & (R'$_2$)SB(s)0/a & --22.02 & 37 & III & bar, arms & dust lane \\
ESO 069-001 & (L)SAB(rl)0$^0$ & --23.32 & 40 & II & bar & \nodata \\
ESO 235-051 & SB(s)0$^+$ & --21.37 & 41.5 & I & bar, arms & blue bulge \\
ESO 265-033 & S0? & --23.44 & 40 & II & boxy bulge, ansae & inclined dust disk \\
ESO 269-013 & (R)SB(r)0$^+$ & --24.01 & 42 & III & bar, rings & two blue rings \\
ESO 274-017 & S0/a(LEDA) & --22.34 & 27 & III & stars projected? & \nodata \\
ESO 316-013 & S0 & --23.62 & 27 & I & bar & central red semiring \\
ESO 324-029 & SAB(r)0$^+$ & --23.47 & 72 & III-II & broad inner ring & inner blue ring \\
ESO 446-049 & SA(r)0$^+$: & --23.15 & 42 & III & outer ring & central oval red ring \\
ESO 469-006 & S0 & --21.21 & 27 & III & \nodata & dust lane \\
ESO 486-038 & S0 & --22.60 & 33 & III & \nodata & blue star projected \\
ESO 496-003 & S0 & --22.80 & 34.5 & I & broad outer ring & red nucleus \\
ESO 506-011 & S0 & --22.80 & 39 & I & \nodata & red nucleus \\
ESO 508-033 & S0? & --22.38 & 24 & III & \nodata & blue nucleus \\
ESO 545-040 & SA(rs)0$^0$? & --20.94 & 47.5 & III & boxy bulge & blue semiring, red nucleus \\
ESO 563-024 & S0 & --23.58 & 34 & I & \nodata & red nucleus \\
ESO 603-029 & (L)SAB0$^0$ & --21.92 & 31 & I & broad ring & asym. dust lane \\
IC 276 & S0$^0$ pec & --22.69 &  50 & III & \nodata & red nucleus, blue bulge \\
IC 537 & (R)S0/a? & --23.83 & 39.5 & III & bar, outer ring & inner red semiring \\
IC 4913 & SA0$-$ & --23.38 & 48 & I & \nodata &  red nucleus \\
NGC 270 & S0$^+$ & --23.29 & 53 & III & shells &  asym. red arm and dust lane \\
NGC 324 & S0(LEDA) & --23.06 & 44 & I & \nodata & blue bulge, red nucleus \\
NGC 1656 & S0$^+$pec: & --23.39 & 52 & III & bar & red thin bar, blue bulge \\
NGC 4087 & SA0$^-$: & --24.30 & 60 & I & \nodata & red nucleus \\
NGC 4878 & SB(r)0$^+$ & --23.98 & 45 & II & bar, outer ring & \nodata \\
NGC 5890 & SA(rl)0$^0$ & --22.44 & 46.5 & I & stellar arms & blue nucleus, red dust arc \\
NGC 6014 & S0 & --23.90 & 52 & II & shells & blue nucleus, red patches \\
NGC 7007 & SA0$^-$ & --23.30 & 102 & III & small bar & east-west color asymmetry \\
NGC 7208 & SAB0$^0$? & --21.64 & 27 & III & stellar arms, merger & blue embedded object \\
PGC 11756 & E(LEDA) & --22.23 & 22 & II & ansae & inner blue ring \\
PGC 16688 & S0(LEDA) & --21.70 & 27 & III & bar & asym. dust lane \\
PGC 34728 & SA0$^-$pec & --22.51 & 50 & III & \nodata & \nodata \\
PGC 35771 & S0/a(LEDA) & --23.25 & 30 & II & bar, arms & red spirals, blue inclusion \\
PGC 46474 & S0/a(LEDA) & --21.04 & 35 & III & ring & blue patchy center \\
PGC 52002 & S0(LEDA) & --22.53 & 29 & I & \nodata & red nucleus \\
PGC 58114 & S0 & --21.16 & 31(K) & III & bar & asym. dust lane \\
PGC 63536 & SA(r)0$^-$ & --21.30 & 42 & III & \nodata & red nucleus \\
PGC 68401 & (R')SB(rs)0$^+$  & --21.91 & 60 & III & ring & asym. blue outer ring/tail \\
UGC 3097 & S0 & --21.71 & 24 & III & bar & blue disk, asym. dust lane \\
UGC 5745 & SB(rs)0$^+$ & --21.54 & 36 & I & bar, ring & asym. dust lane, red arms \\
\enddata
\tablenotetext{a}{Mostly from NED; but some data taken from HyperLEDA \citep{hyperleda} are marked by `(LEDA)'.}
\tablenotetext{b}{Mostly the optical radii are taken from HyperLEDA \citep{hyperleda}; but some NIR radii taken from NED
are marked by `(K)'.}
\end{deluxetable}
}

The relative thicknesses of the disks characterizing the ratio of the vertical and radial scalelengths
were calculated by our novel method described in details by \citet{thickmeth}.
The careful testing of the method and determination of the boundaries of its applicability can be found in
\citet{chudpaper}.  In a few words, we used the analysis of the projection effects for an oblate ellipsoid
with the axes $a_1=a_2>a_3$ made by \citet{hubble26}. If we look at the intrinsically round, infinitely thin disk
projected onto the sky plane under the inclination $i$ (here we fix $i=90$\degr\ for edge-on disks), we would see an ellipse
with the axis ratio of $b/a=\cos i$. If the disk is not infinitely thin and can be described by an oblate ellipsoid
with the vertical-to-radial axis ratio of $q$, then $\cos ^2 i =\frac{(b/a)^2 - q^2}{1-q^2}$. The latter equation
enables us to calculate the relative disk thickness $q$, if we invent a possibility to measure independently
the inclination $i$ and the isophote axis ratio $b/a$. The latter parameter, a visible axis ratio, is provided by the isophote
analysis. As for the former parameter, we can obtain it directly from the 2D surface photometry if we assume that the disk under consideration
has an exponential radial brightness profile. Indeed, the exponential scalelength for a given galactic disk can be used as a standard rule,
characterizing projection effects. Its measurements, taken under different azimuth angles, would follow a pure cosine law  and
would vary from $h$ along the major axis to $h \cos i$ along the minor axis. The knowledge of the 'projected h' azimuthal changes
provides an independent estimate of the inclination $i$. We split commonly a full
galaxy image into 20 sectors, with 18-degree opening angle each, and measure surface-brightness profiles within
each of them. These surface-brightness profiles are fitted by exponentials, and the on-plane azimuthal distribution
of the 20 (projected) scalelengths obtained in such a way is approximated by an ellipse. Just this scalelength ellipse
must have an axis ratio equal to $\cos i$ so giving us a possibility to determine the disk inclination. Then, having in hands
the disk inclination $i$ and the isophote ellipticity, we can estimate of the disk relative thickness. Namely, by measuring the isophote
ellipticity $e_I$ and the ellipticity of the azimuthal distribution of the projected exponential scalelengths $e_h$, we immediately
derive the exponential disk relative thickness from the following expression:

\begin{equation}
q = \sqrt{1 - \frac{2e_I - e_I ^2}{2e_h - e_h ^2}}.
\label{finform}
\end{equation}
.

\section{The results and discussion}

\subsection{Variety of the isolated S0s}

\begin{figure*}[bpt!]
\centering
\begin{tabular}{c c c c}
 \includegraphics[width=4cm]{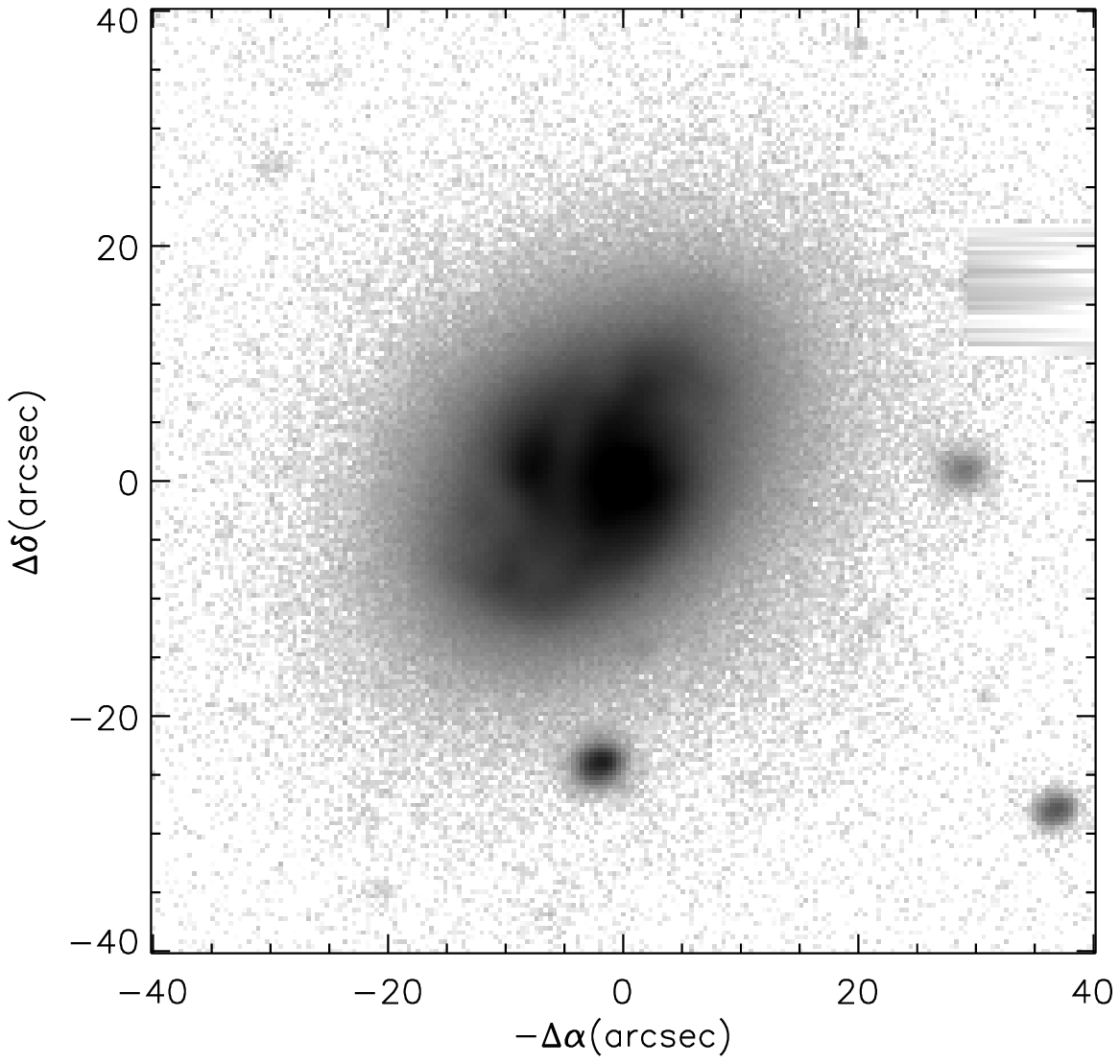} & \includegraphics[width=4cm]{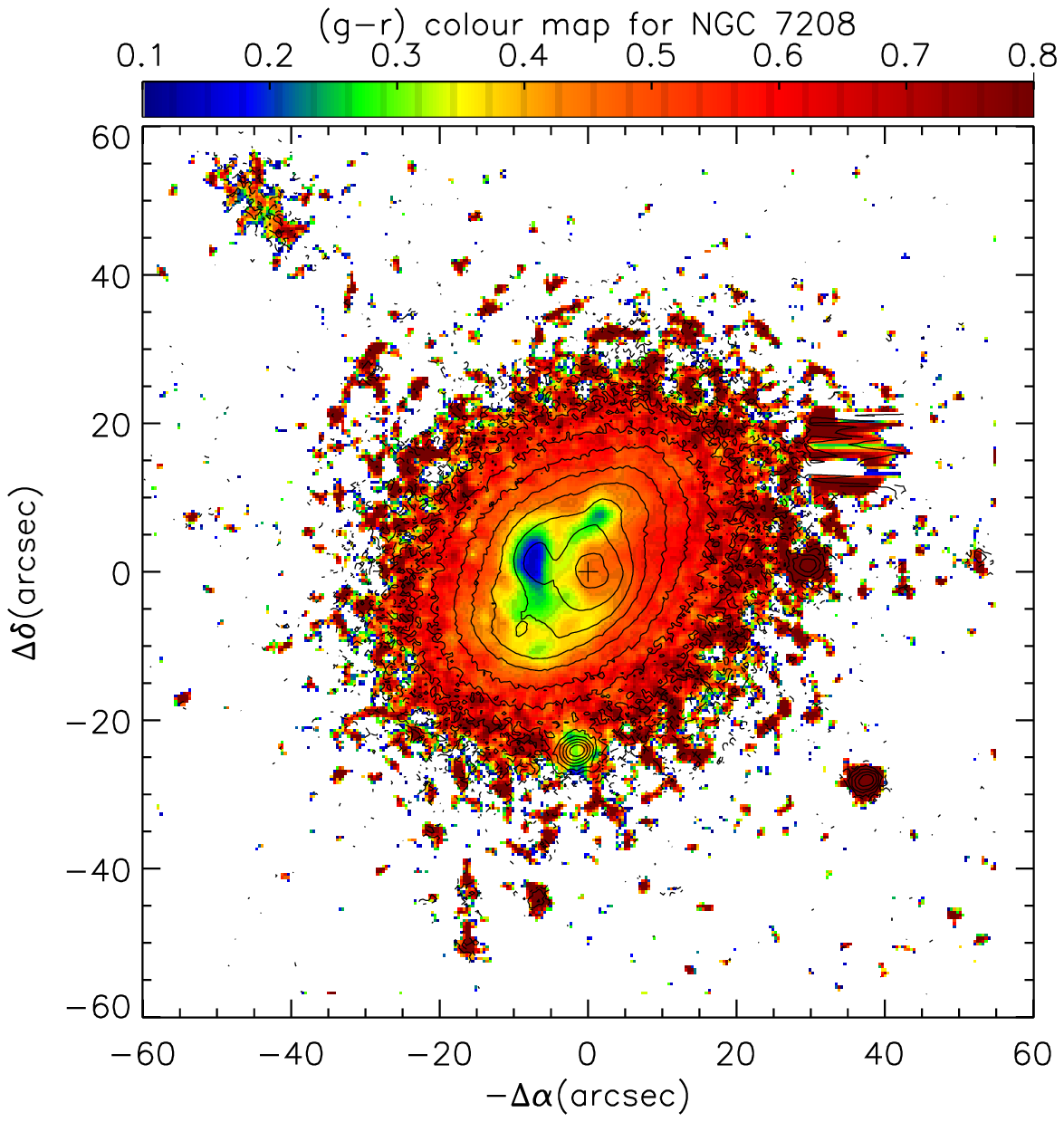} &
 \includegraphics[width=4cm]{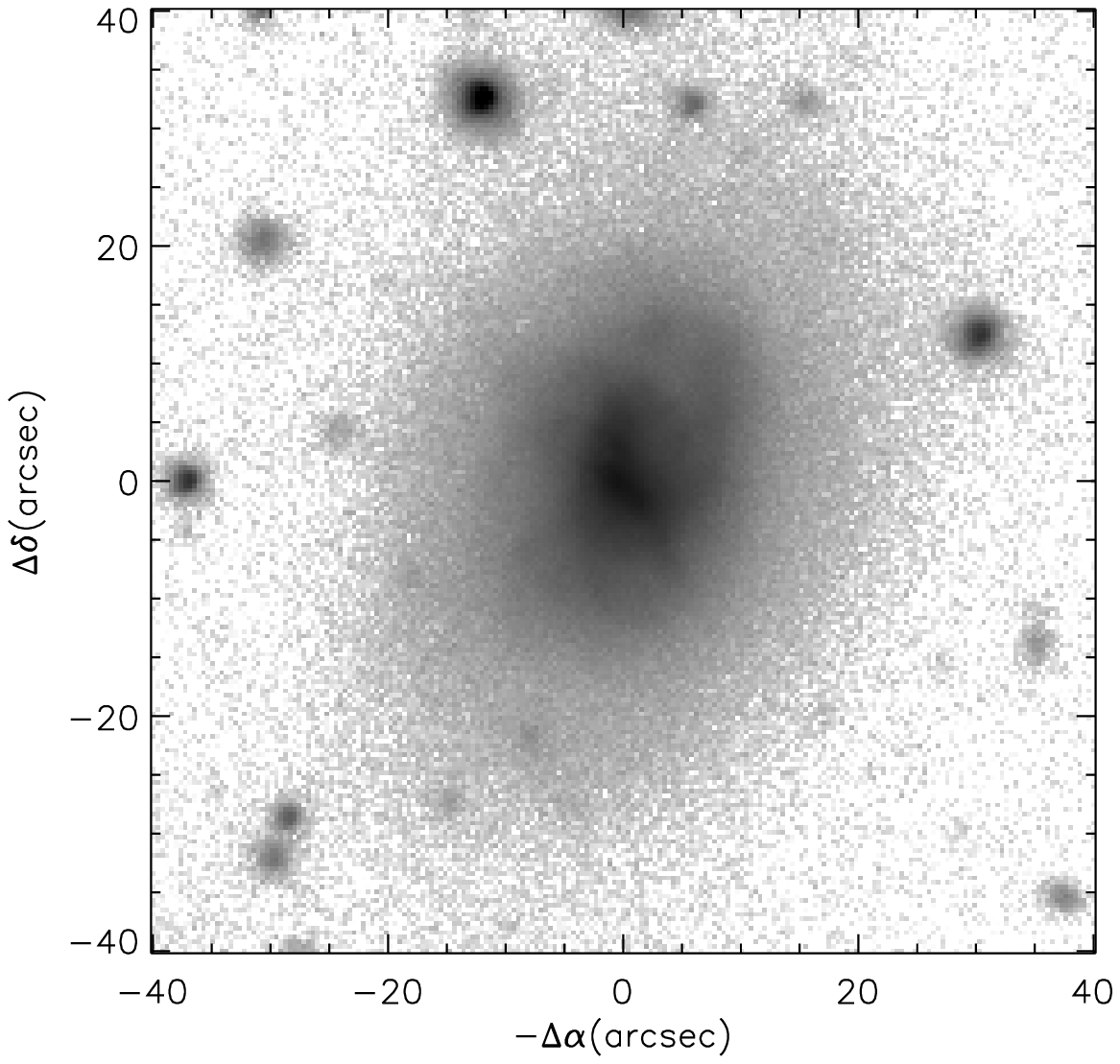} &
 \includegraphics[width=4cm]{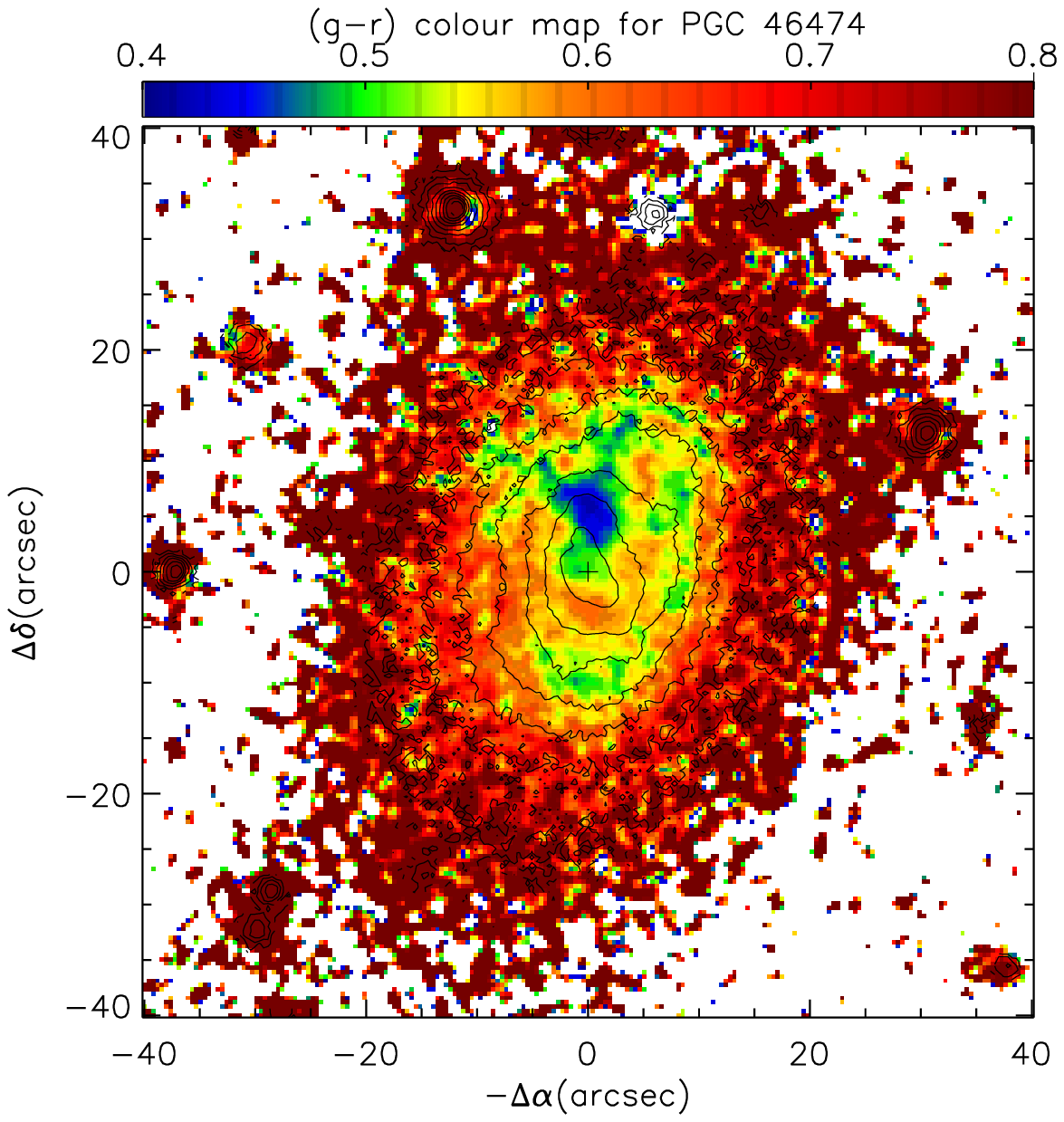} \\
\includegraphics[width=4cm]{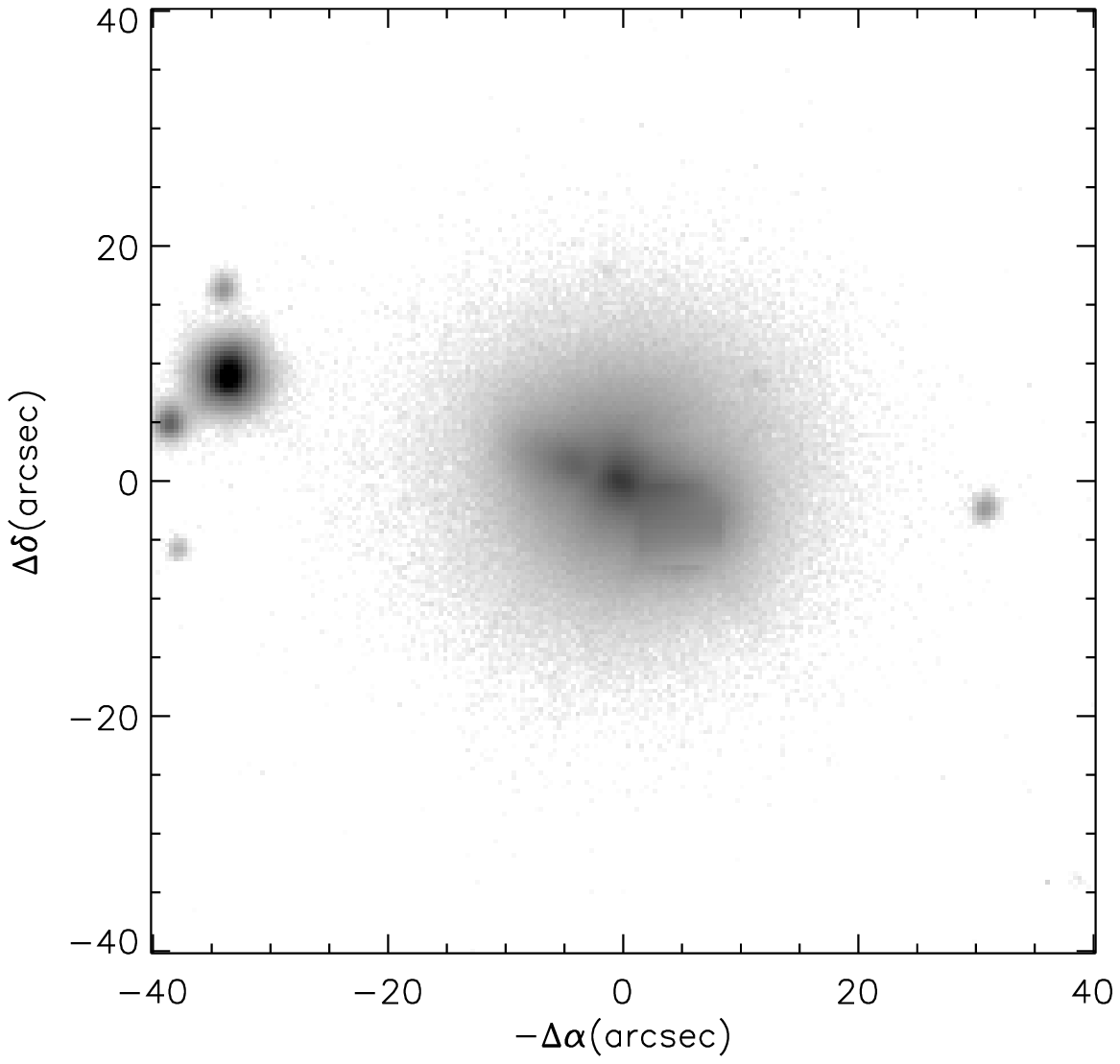} & \includegraphics[width=4cm]{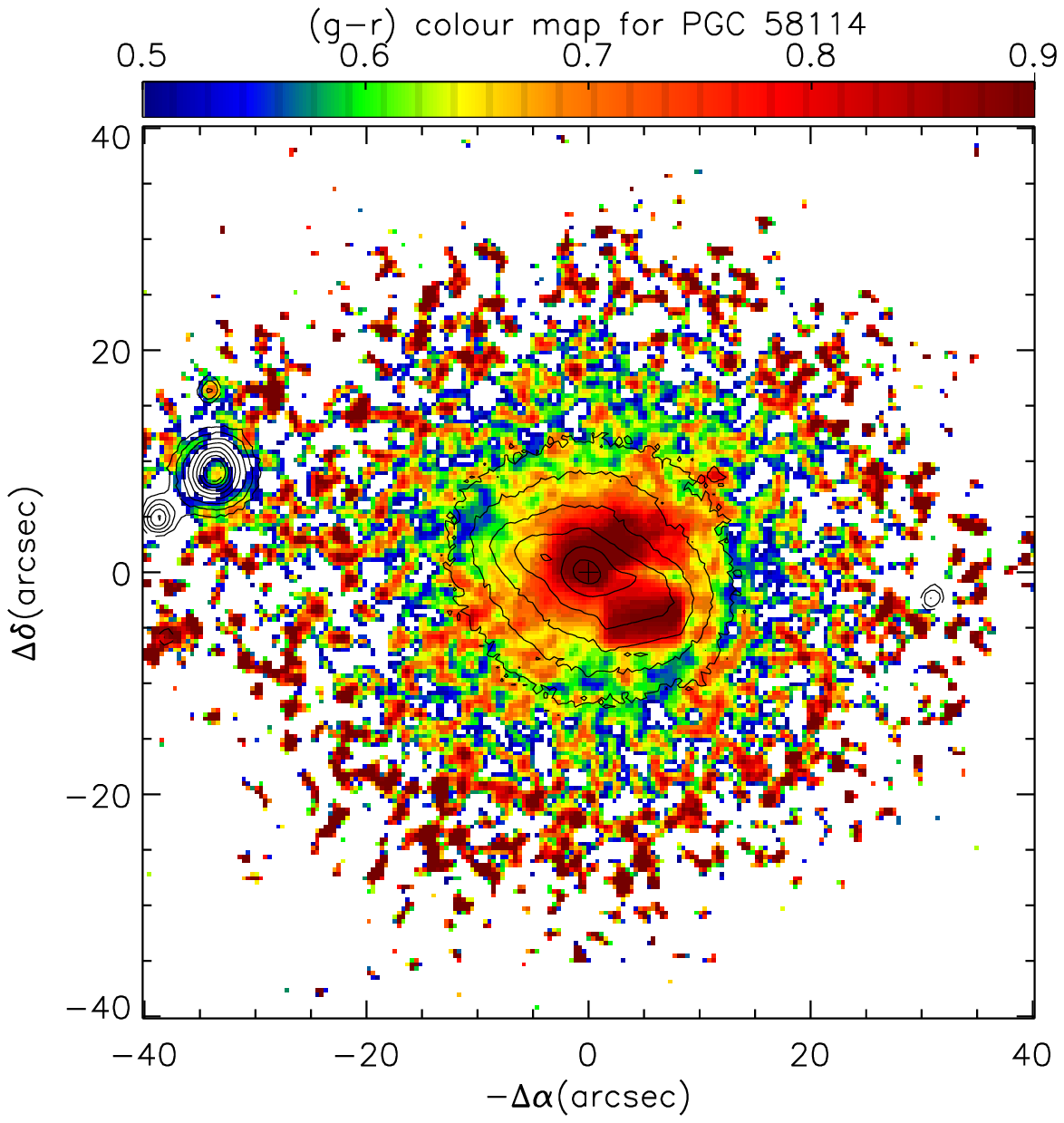} &
\includegraphics[width=4cm]{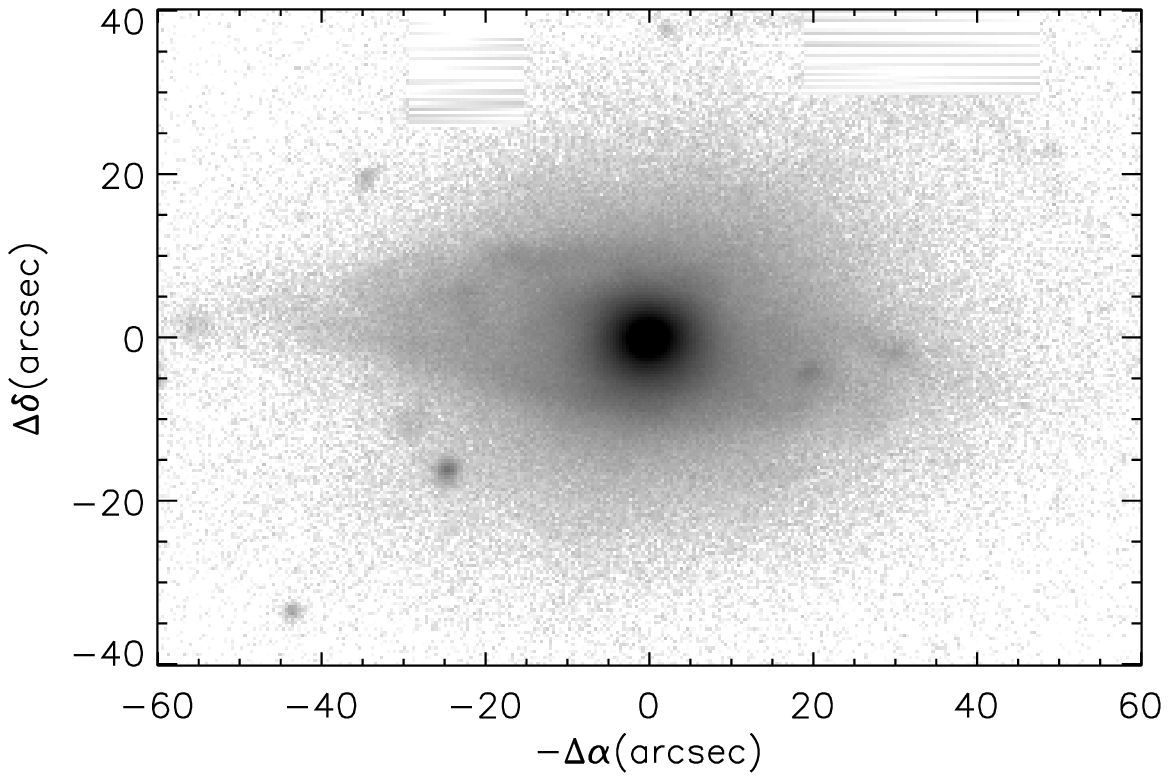} &
 \includegraphics[width=4cm]{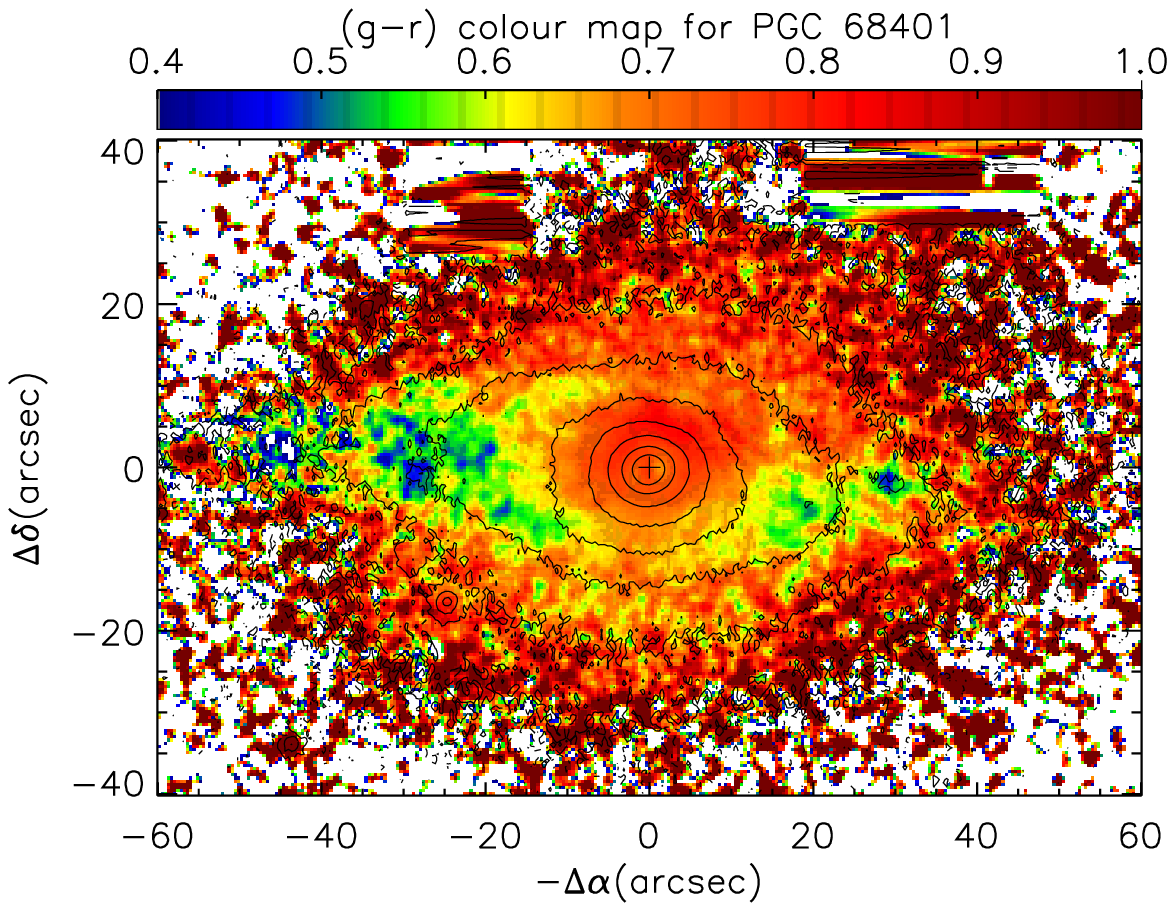} \\
\end{tabular}
\caption{Some examples of the merger signatures in the dwarf isolated S0s. In every pair of plots
the left one shows the $r$-band map, in logarithmic flux scale, and the right one -- the color map,
of $g-r$.}
\label{mergers}
\end{figure*}

\begin{figure*}[bpt!]
\centering
\begin{tabular}{c c}
 \includegraphics[width=7cm]{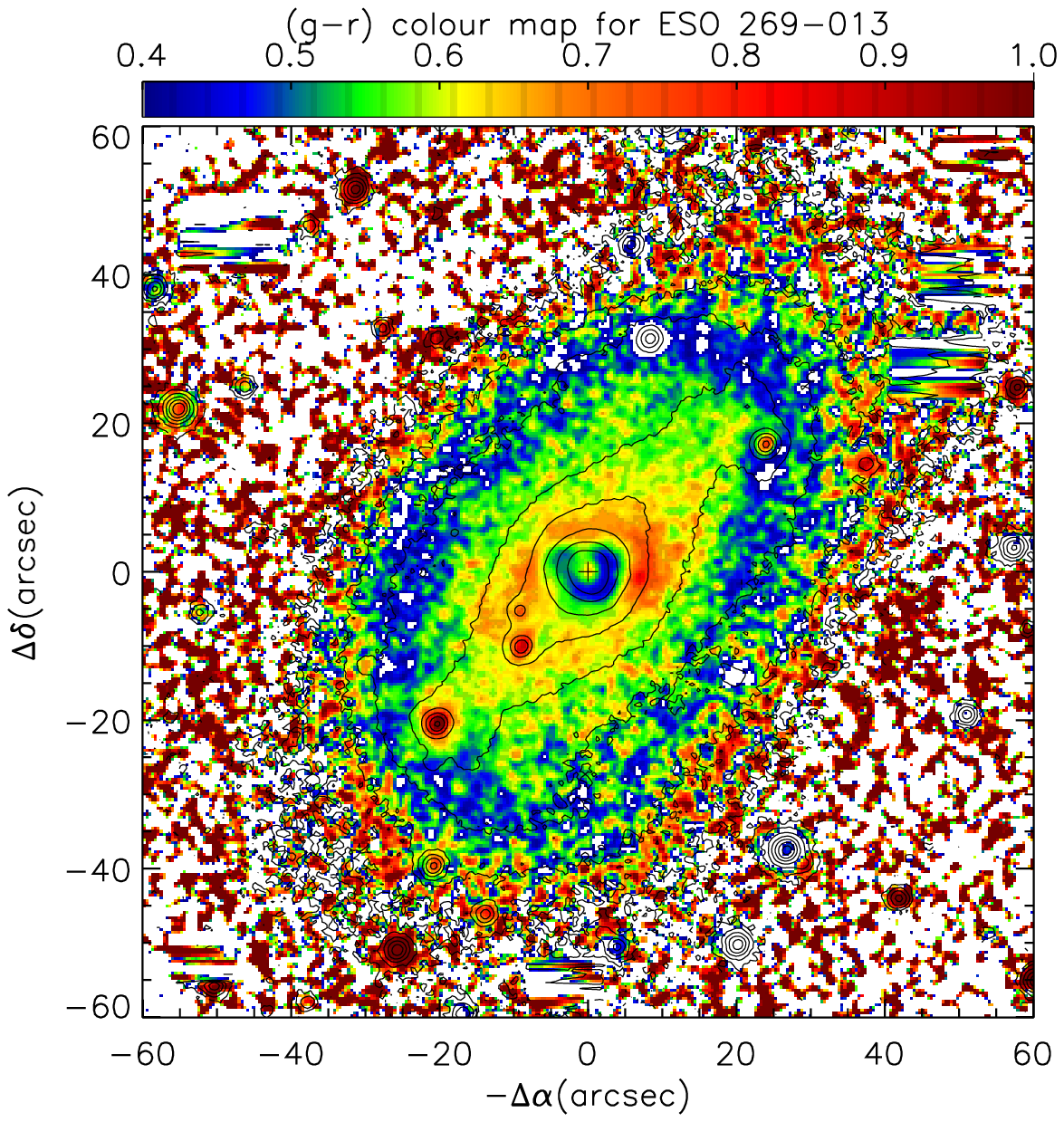} &
 \includegraphics[width=7cm]{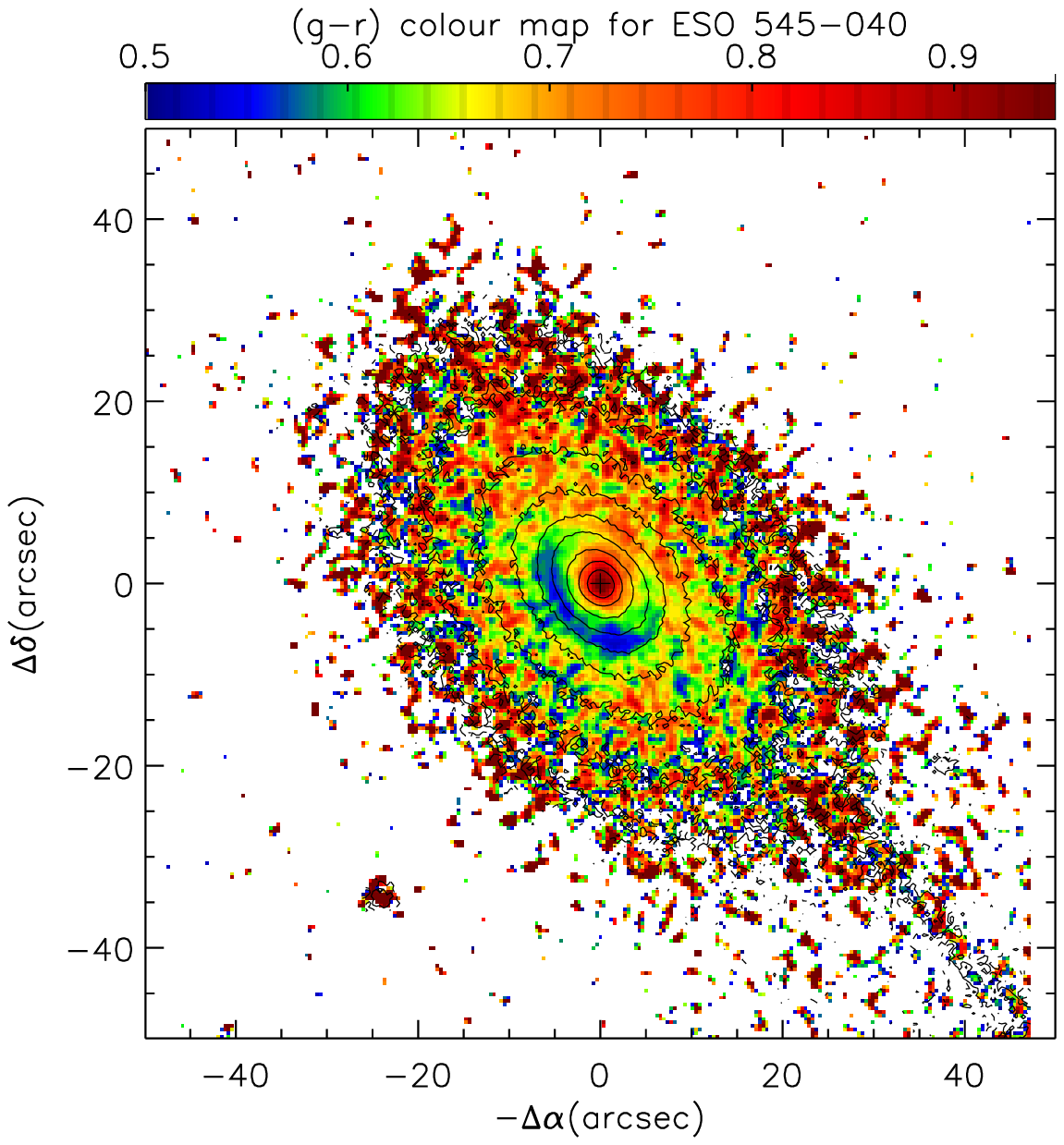} \\
\includegraphics[width=7cm]{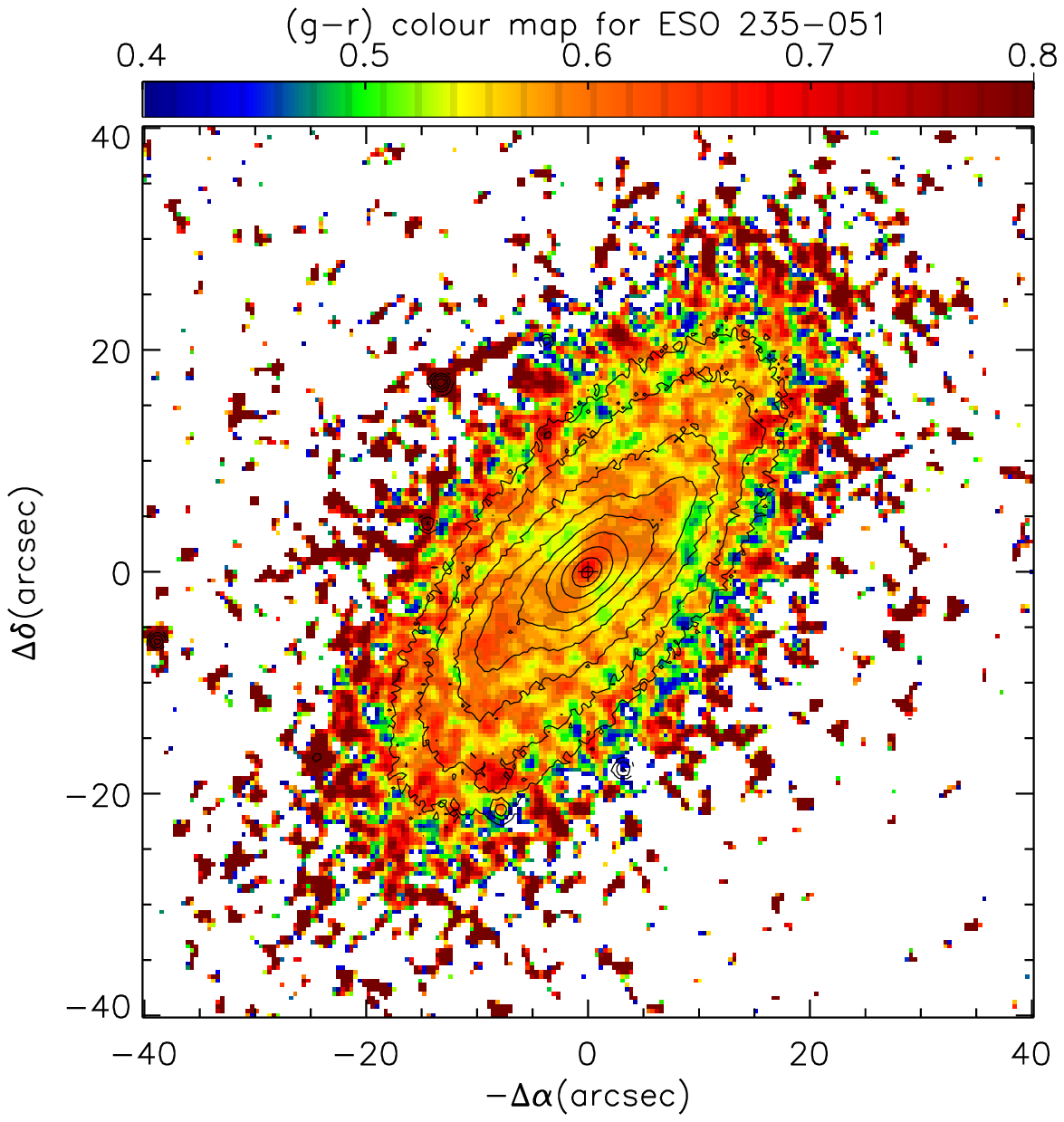} &
 \includegraphics[width=7cm]{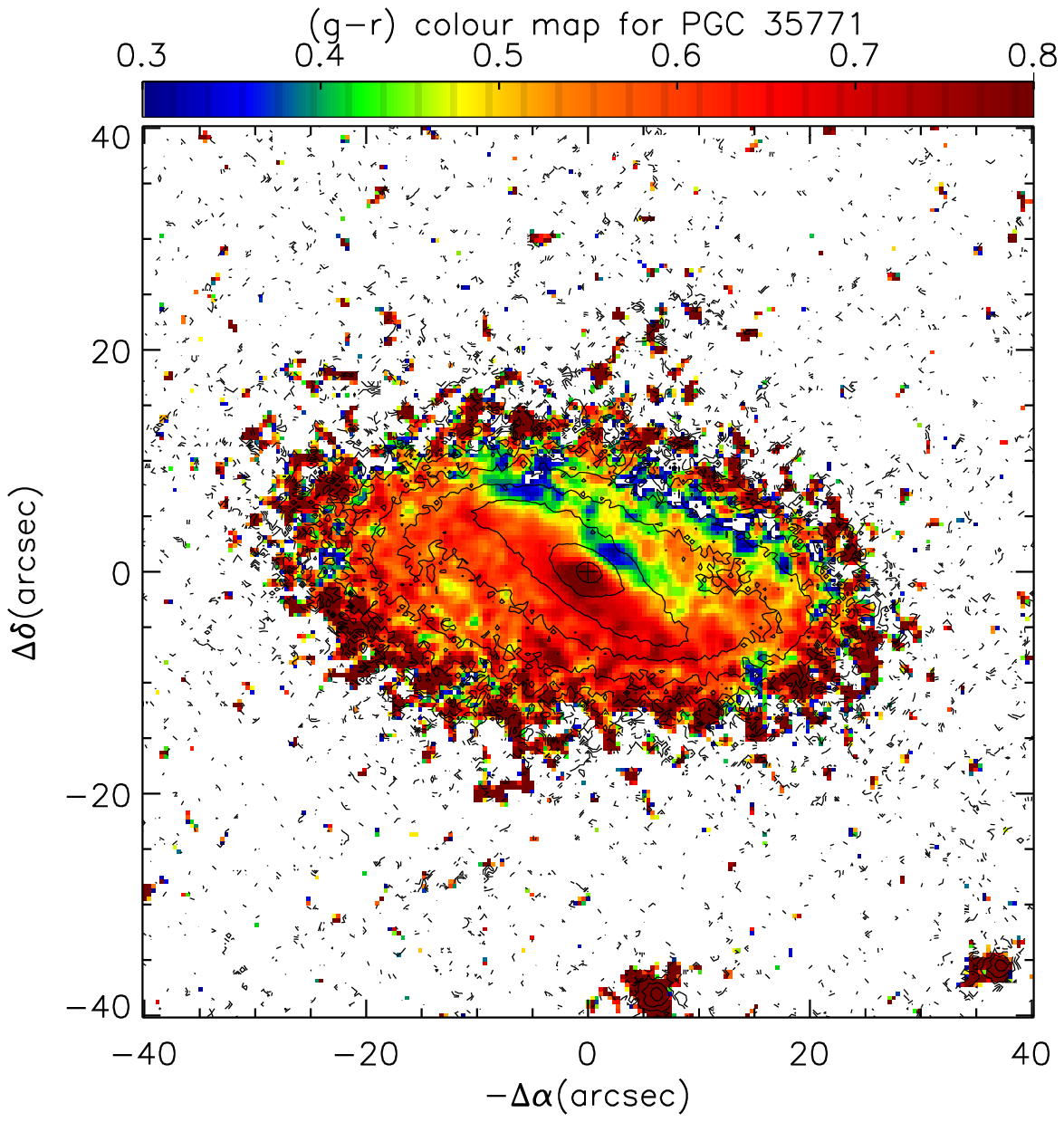} \\
 \includegraphics[width=7cm]{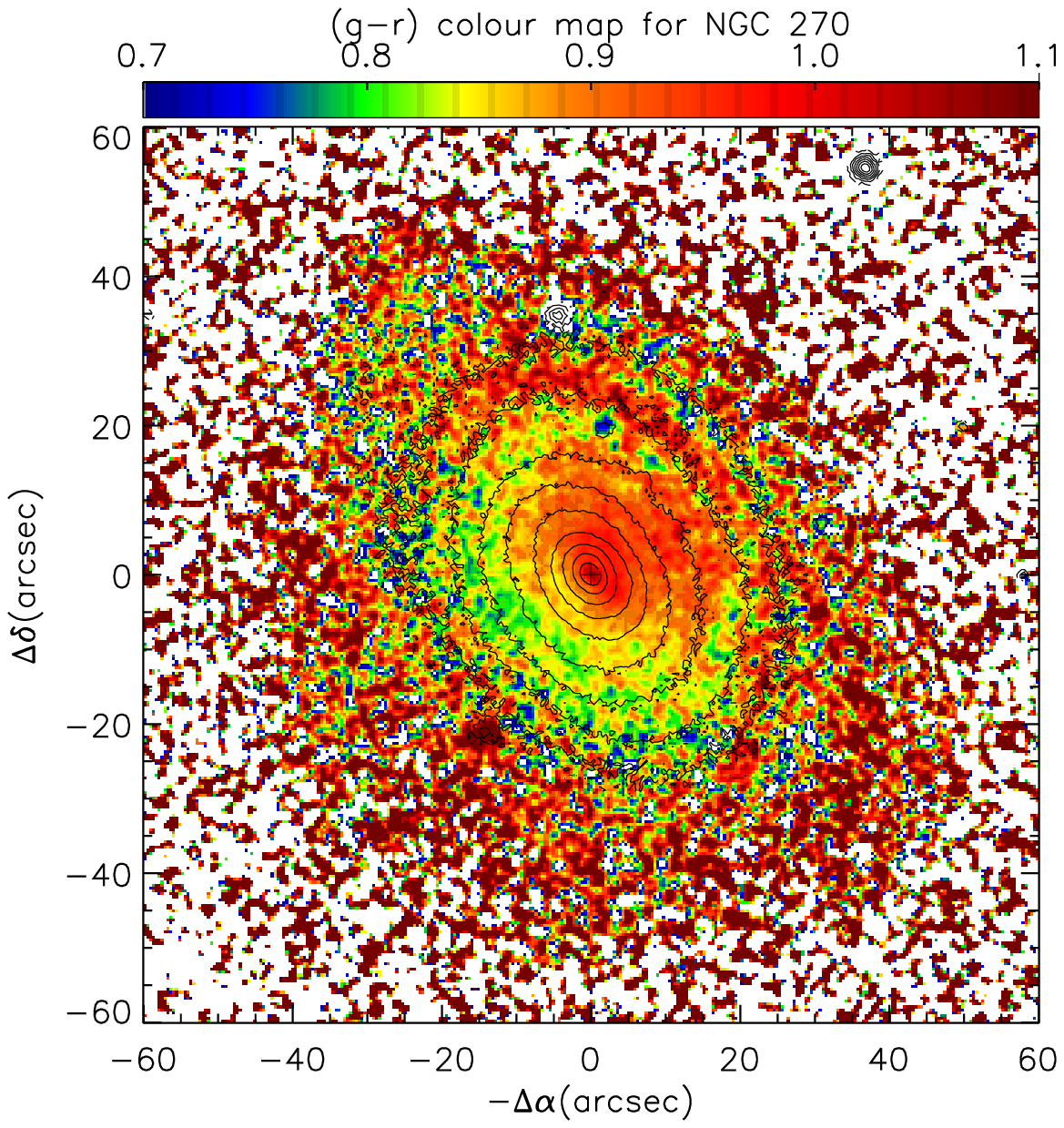} &
 \includegraphics[width=7cm]{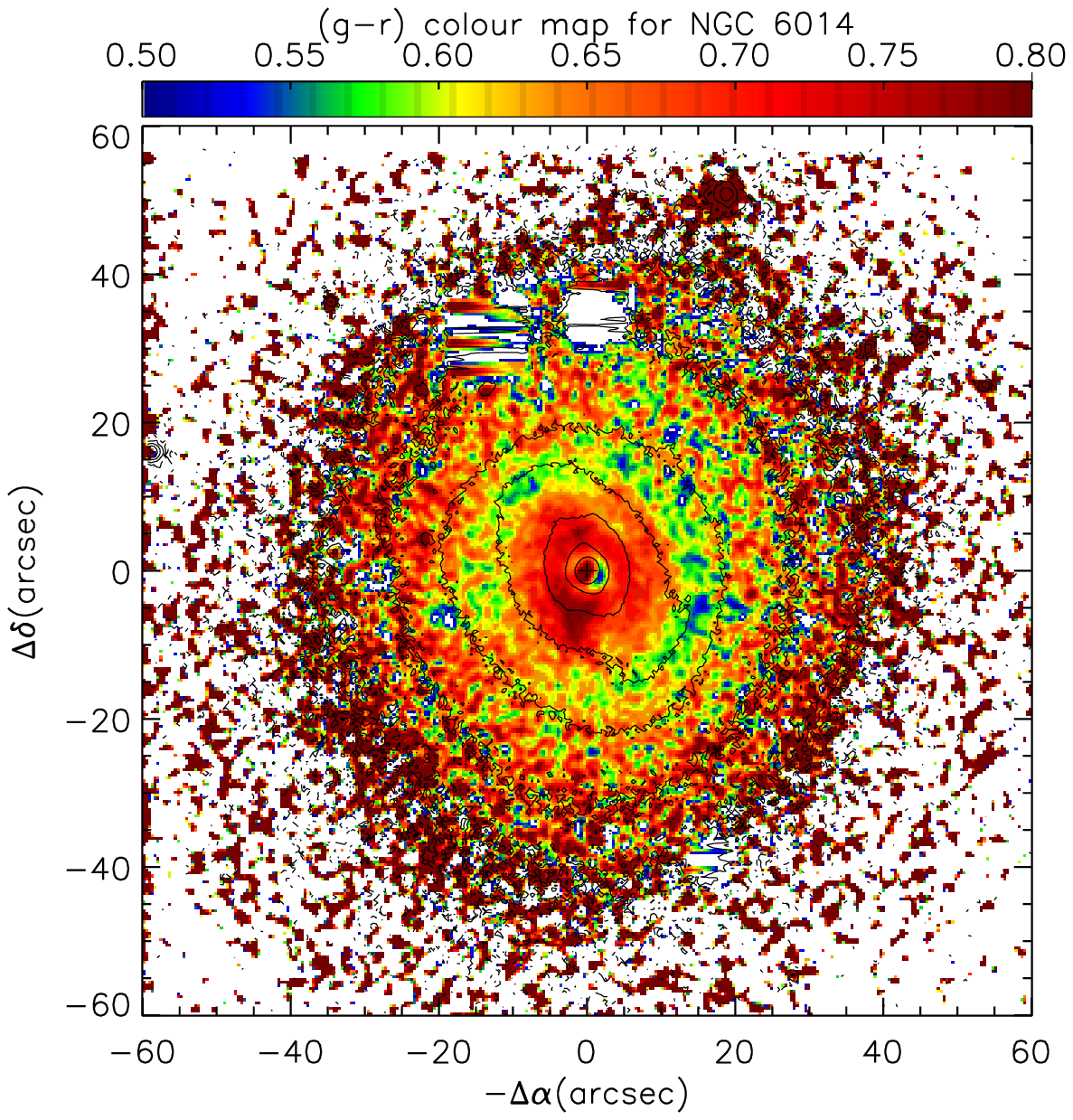} \\
\end{tabular}
\caption{Some examples of the color structures: {\it upper row} -- circumnuclear and inner blue rings, {\it middle
row} -- red nuclear spiral arms, {\it bottom row} -- shells, both reddish or blue. The isophotes overposed represent
the $r$-band surface brightness levels.}
\label{colormap}
\end{figure*}

Table~\ref{listgal} lists all 42 galaxies which we have analyzed, with specifying also the type of the disk
surface-brightness profile which we have determined, and with attached notes about additional structure details and color features
seen in the $g-r$ maps. The presence of a bar was recognized by us through visual morphological classification,
or when the isophote ellipticity profile had a distinct local maximum exceeding the outer ellipticity level related to the disk
spatial orientation. The existence of a ring is fixed through the visual inspection of the surface-brightness profiles; the rings
can be or not to be highlighted by the color. The color distributions demonstrate many small-scale features: 
just isolated S0s are not so smooth as we are in the habit of thinking. The low-luminosity S0s possess sometimes 
blue nuclei; moreover, the low-luminosity isolated S0s are often looking like merger remnants (Fig.~\ref{mergers}). 
Among the galaxies of all luminosities the blue rings of various sizes or circumnuclear red (dust) lanes can also be noted.
We show some typical cases of the inhomogeneous color distributions in our galaxies in Fig.~\ref{colormap}.

We give the results of our analysis for the radial and vertical structures of the $r$-band images of the galactic disks
in the Table~\ref{s0type1} for the brightness profiles of Type I, in the Table~\ref{s0type2} for the
brightness profiles of Type II, and in the Table~\ref{s0type3} for the brightness profiles of Type III, correspondingly.
The radial surface-brightness profile parameters -- central surface brightnesses $\mu _0$ and exponential
scalelengths $h$ -- are calculated by fitting the profiles with exponential functions in the radius ranges noted in
the Tables. Following \citet{freeman}, we tried to use the radius ranges which were at least twice larger
than the fitted exponential scalelengths, but for the Type II profiles the inner segments were often shorter, being of
order of one exponential scalelengths. The $\mu _0$'s given in the Tables are not corrected for the intrinsic dust;
only Galactic extinction provided by NED is taken into account.
The relative thicknesses of the stellar disks, $q$, are also given in the Tables; they have been calculated
only for the high signal-to-noise images, not contaminated by foreground stars, and which are also far 
from edge-on or face-on orientation. The relative thicknesses $q$ of the stellar disks, characterizing the ratio
of the vertical and radial scalelengths, are presented in the Table~\ref{s0type1} for the entire single-scale
exponential disks, but in the Table~\ref{s0type2} and Table~\ref{s0type3} only for the inner segments of the truncated and antitruncated disks.

\begin{table*}
\caption{Parameters of the Type-I disks}
\label{s0type1}
\begin{center}
\begin{tabular}{lccrrc}
\hline\noalign{\smallskip}
Galaxy & Radius range, $^{\prime \prime}$ & $\mu _0$(r) & $h_r$, $^{\prime \prime}$ 
& $h_r$, kpc & q \\
\hline
ESO 235-051 & 10--40 & 19.2 & $8.6\pm 0.0$ & $1.2\pm 0.0$ & $0.298 \pm 0.006$ \\
ESO 316-013 & 20--40 & 19.2 & $10.9\pm 0.01$ & $2.8\pm 0.0$ &   \\
ESO 496-003 & 8--25 & 18.8 & $8.0\pm 0.0$ & $1.4\pm 0.0$ & $0.536 \pm 0.001$ \\
ESO 506-011 & 18--45 & 20.1 & $9.0\pm 0.1$ & $2.1\pm 0.0$ &  \\
ESO 563-024 & 20--50 & 20.3 & $13.1 \pm 0.0$ & $2.6\pm 0.0$ &  \\
ESO 603-029 & 8--33 & 18.5 & $6.0\pm 0.05$ & $1.20\pm 0.01$ &  $0.303 \pm 0.011$ \\
IC 4913 & 10--43 &  20.0 &  $11.23\pm 0.01$ & $2.5\pm 0.0$ & $0.40 \pm 0.02$ \\
NGC 324 & 20--60 & 19.3 & $9.9\pm 0.0$ & $2.1\pm 0.0$ & $0.224 \pm 0.043$ \\
NGC 4087 & 25--75 & 19.7 & $18.0\pm 0.0$ & $4.3\pm 0.0$ & $0.548 \pm 0.026$ \\
NGC 5890 & 10--60 & 19.3 & $12.48\pm 0.01$ & $1.9 \pm 0.0$ & $0.47 \pm 0.02$ \\
PGC 52002 & 15--48 & 19.9 & $8.62\pm 0.01$ & $1.6\pm 0.0$ & $0.190 \pm 0.015$ \\
UGC 5745 & 10--50 & 19.1 & $8.05\pm 0.00$ & $0.7\pm 0.0$ &    \\
\hline
\end{tabular}
\end{center}
\end{table*}

\begin{table*}
\caption{Parameters of the Type-II disks}
\label{s0type2}
\begin{flushleft}
\scriptsize
\begin{tabular}{lccrrccccrr}
\hline
Galaxy & \multicolumn{5}{c}{Inner disk} &  & \multicolumn{4}{c}{Outer disk} \\
 & Radius range, $^{\prime \prime}$ & $\mu _0$(r) & $h_r$, $^{\prime \prime}$ 
& $h_r$, kpc & q & $\mu _{brk}$ & Radius range, $^{\prime \prime}$ & $\mu _0$(r) & $h_r$, $^{\prime \prime}$
& $h_r$, kpc  \\
\hline
ESO 069-001 & 12--28 & 19.1:: & $8.23\pm 0.16$ & $1.93\pm 0.04$ & & 22.4:: & 27--34  & 17.1:: & $5.27\pm 0.06$ & $1.24\pm 0.01$  \\
ESO 265-033 & 9--19 & 19.3 & $9.1\pm 0.2$ & $2.65\pm 0.06$ & 0.18 & 21.64 & 20--34 & 18.3 & $6.39\pm 0.14$ & $1.86\pm 0.04$  \\
ESO 324-029 & 40--60 & 19.7 & $23.76\pm 0.03$ & $3.82\pm 0.00$ & 0.23 & 22.6 & 60--105 & 18.0 & $15.02\pm 0.02$ & $2.42\pm 0.00$ \\
NGC 4878 & 16--30 & 21.0 & $29.51\pm 0.06$ & $8.03\pm 0.02$ &  & 22.2 & 37--67 & 19.1 & $11.36\pm 0.13$ & $ 3.09\pm 0.04$ \\
NGC 6014 & 22--34 & 20.5  & $16.31\pm 0.01$ & $2.67\pm 0.00$ & 0.59 & 23.1 & 35--65 & 19.8 & $12.88\pm 0.01$ & $2.1\pm 0.0$ \\
PGC 11756 & 1--8 & 18.7 &$7.8\pm 0.15$ & $2.25\pm 0.04$ &  & 19.8 & 10--25 & 17.6 & $3.77\pm 0.03$ & $1.1\pm 0.0$ \\
PGC 35771 & 6--16 & 19.9 & $9.0\pm 0.2$ & $2.66\pm 0.06$ & 0.33 & 21.0 & 20--32 & 19.5 & $6.54\pm 0.31$ & $1.95\pm 0.09$ \\
\hline
\end{tabular}
\end{flushleft}
\end{table*}

\begin{table*}
\caption{Parameters of the Type-III disks}
\label{s0type3}
\begin{flushleft}
\scriptsize
\begin{tabular}{lccrrccccrr}
\hline
Galaxy & \multicolumn{5}{c}{Inner disk} &   & \multicolumn{4}{c}{Outer disk} \\
 & Radius, $^{\prime \prime}$ & $\mu _0$(r) & $h_r$, $^{\prime \prime}$
& $h_r$, kpc & q & $\mu _{brk}$ & Radius, $^{\prime \prime}$ & $\mu _0$(r) & $h_r$, $^{\prime \prime}$
& $h_r$, kpc  \\
\hline
ESO 003-001 & 11--30 & 19.1 & $9.0\pm 0.0$ & $2.4\pm 0.0$ &  & 22.9 & 30--47 & 20.7 & $15.3\pm 0.6$ & $4.05\pm 0.16$ \\
ESO 040-002 & 10--28 & 19.8(g):: & $10.5\pm 0.0$ & $2.1\pm 0.0$ & $0.61\pm 0.02$ & 22.6(g):: & 30--48 & 20.8(g):: & $16.4\pm 0.4$ & $3.26\pm 0.08$ \\
ESO 052-014 & 9--27 & 19.1 & $8.2\pm 0.1$ & $1.6\pm 0.0$ &  & 22.8 & 27--50 & 20.8 & $15.2\pm 0.4$ & $3.0\pm 0.1$ \\
ESO 269-013 & 15--50 & 20.2 & $16.16\pm 0.02$ & $4.28\pm 0.01$ & $0.53\pm 0.01$ & 23.8 & 50--85 & 21.1 & $21.5\pm 0.5$ & $5.7\pm 0.1$ \\
ESO 274-017 & 5--11 & 18.6 & $3.8\pm 0.08$ & $0.87\pm 0.02$ &   & 21.9 & 12--22 & 20.3 & $7.8\pm 0.3$ & $1.78\pm 0.07$ \\
ESO 446-049 & 6--17 & 18.7 & $6.4\pm 0.05$ & $1.7\pm 0.0$ & $0.43\pm 0.02$ & 21.15 & 20--55 & 19.6 & $10.15\pm 0.005$ & $2.70\pm 0.00$ \\
ESO 469-006 & 5--13 & 19.0 & $4.11\pm 0.04$ & $0.73\pm 0.01$ & $0.105\pm  0.011$ & 22.5 & 15--32 & 20.0 & $6.0\pm 0.1$ & $1.07\pm 0.02$ \\
ESO 486-038 & 15--40 & 20.5 & $11.1\pm 0.0$ & $3.0\pm 0.0$ &   & 24.8 & 42--57 & 22.3 & $19\pm 3$ & $5.2\pm 0.8$ \\
ESO 508-033 & 4--13 & 18.0 & $3.95\pm 0.03$ & $0.94\pm 0.01$ & $0.22\pm 0.01$ & 21.9 & 16--28 & 19.6 & $6.54\pm 0.01$ & $1.55\pm 0.00$ \\
ESO 545-040 & 15--45 & 20.1 & $13.4\pm 0.0$ & $1.1\pm 0.0$ & $0.35\pm 0.01$ & 23.7 & 45--60 &  21.0  & $17.9\pm 1.3$ & $1.5\pm 0.1$ \\
IC 276 & 12--25 & 17.7 & $7.0\pm 0.05$ & $1.26\pm 0.01$ &   & 21.85 & 30--60 & 19.8 & $14.1\pm 0.2$ & $2.54\pm 0.04$ \\
IC 537 & 5--15 & 17.5 & $4.05\pm 0.04$ & $1.10\pm 0.01$ & $0.22\pm 0.02$ & 21.5 & 18--64 & 20.6 & $17.66\pm 0.02$ & $4.82\pm 0.01$ \\
NGC 270 & 12--25 & 19.3 & $10.3\pm 0.1$ & $2.33\pm 0.02$ & $0.46\pm 0.02$ & 22.3  & 35--70 & 20.8 & $20.7\pm 0.0$ & $4.7\pm 0.0$ \\
NGC 1656 & 15--30 & 19.4 & $12.00\pm 0.01$ & $2.9\pm 0.0$ & $0.54\pm 0.02$ & 22.3 & 35--80 & 20.6 & $20.34\pm 0.02$ & $4.96\pm 0.00$ \\
NGC 7007 & 20--40 & 19.2 & $12.57\pm 0.01$ & $2.43\pm 0.00$ & $0.26\pm 0.01$ & 23.0 & 45--75 & 20.8 & $21.9\pm 0.6$ & $4.23\pm 0.12$ \\
NGC 7208 &  9--20 & 18.0 &  $4.50\pm 0.04$  &  $0.70\pm 0.01$ & $0.44\pm 0.02$ & 23.5 & 28--48 & 21.4 & $11.44\pm 0.43$ & $1.78\pm 0.07$ \\
PGC 16688 & 5--12 & 19.1 & $4.36\pm 0.05$ & $0.85\pm 0.01$ & $0.554\pm 0.026$ & 21.9 & 12--24 & 20.0 & $6.5\pm 0.1$ & $1.27\pm 0.02$ \\
PGC 34728 & 9--24 & 20.0 & $8.55\pm 0.15$ & $2.3\pm 0.04$ &   & 22.4 & 28--54 & 20.9 & $13.6\pm 0.6$ & $3.6\pm 0.2$ \\
PGC 46474 & 3--18 & 19.3 & $6.4\pm 0.04$ & $1.22\pm 0.01$ & $0.36\pm 0.01$ & 22.75 & 20--40 & 20.8 & $11.28\pm 0.15$ & $2.14\pm 0.03$ \\
PGC 58114 & 3--14 & 18.1 & $4.40\pm 0.03$ & $0.46\pm 0.00$ &   & 21.8 & 16--30 & 20.2 & $10.3\pm 0.2$ & $1.07\pm 0.02$ \\
PGC 63536 & 13--28 & 19.9 & $10.34\pm 0.16$ & $1.24\pm 0.02$ & $0.41\pm 0.02$ & 22.9 & 36--52 & 21.1 & $17.3\pm 1.6$ & $2.08\pm 0.19$ \\
PGC 68401 & 13--28 & 20.4 & $13.07\pm 0.01$ & $2.10\pm 0.00$ & $0.49\pm 0.02$ & 23.25 & 40--75 & 21.9 & $27.36\pm 0.05$ & $4.40\pm 0.01$ \\
UGC 3097 & 6--16 & 18.8 & $5.12\pm 0.04$ & $1.18\pm 0.01$ & $0.43\pm 0.02$ & 21.65 & 20--33 & 19.4 & $6.5\pm 0.25$ & $1.50\pm 0.06$ \\
\hline
\end{tabular}
\end{flushleft}
\end{table*}

\subsection{Radial structure of the S0 disks in various environments}

\begin{figure*}[t]
\centering
\begin{tabular}{c c c}
 \includegraphics[width=5cm]{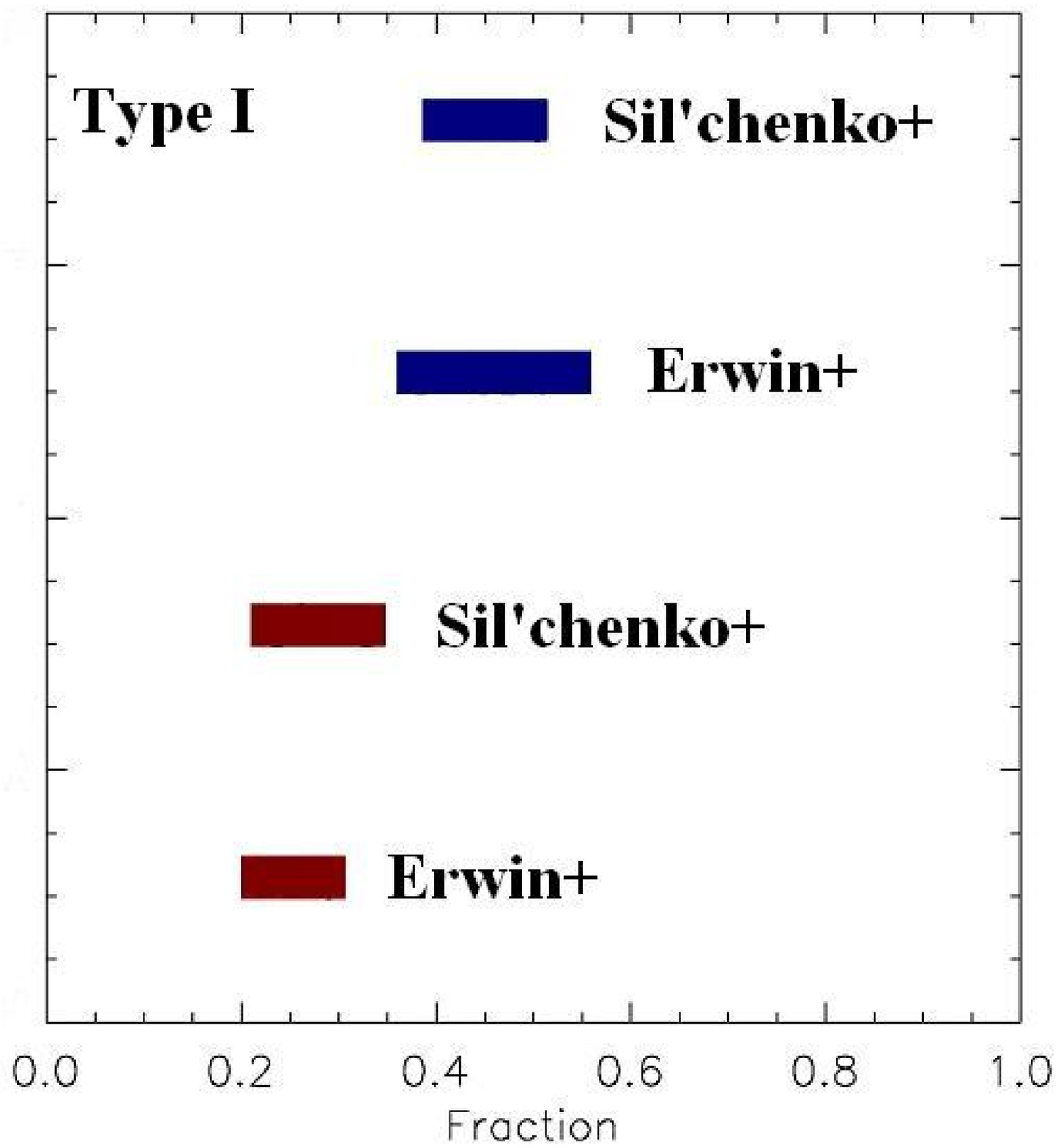} &
 \includegraphics[width=5cm]{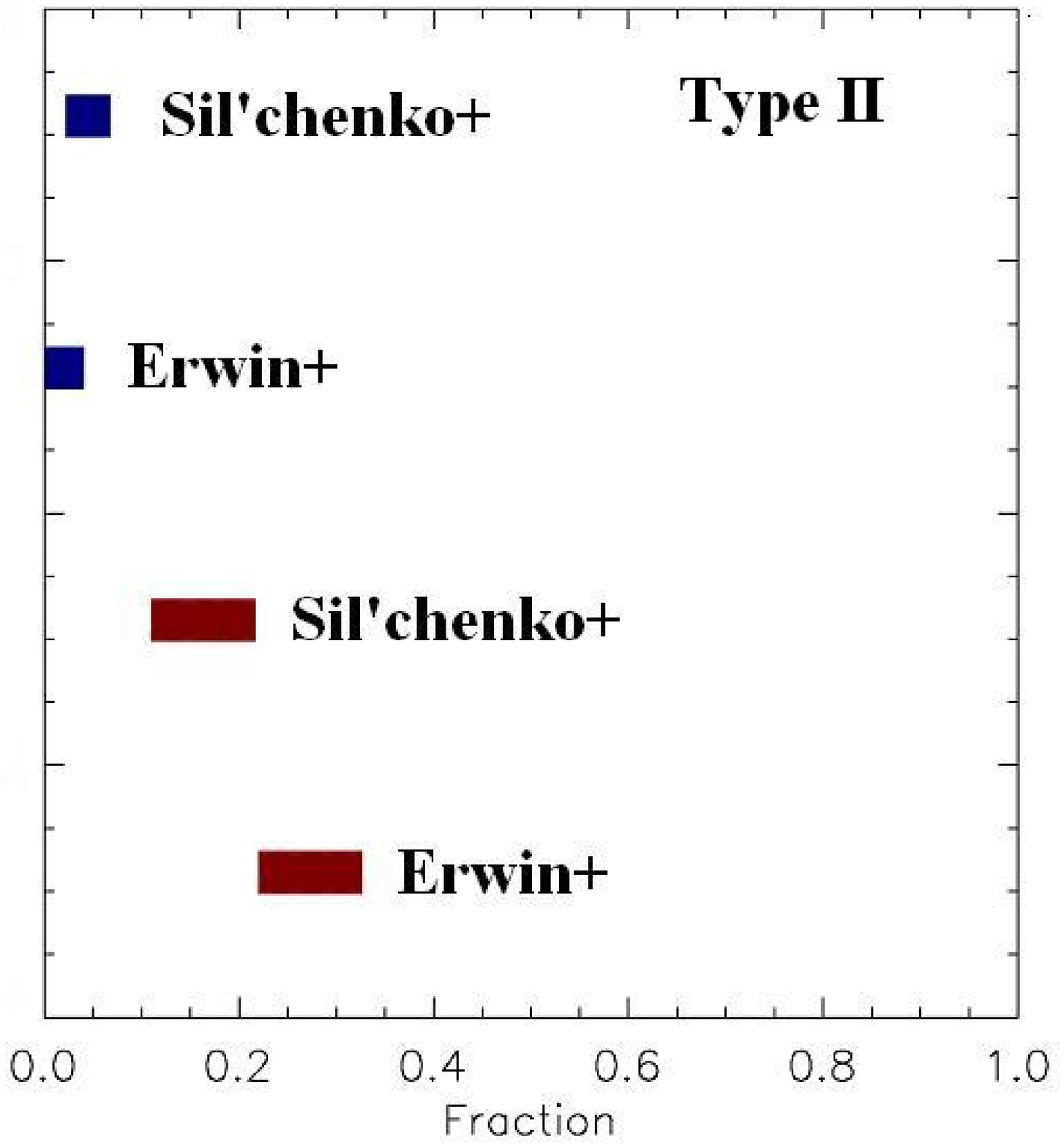}  &
 \includegraphics[width=5cm]{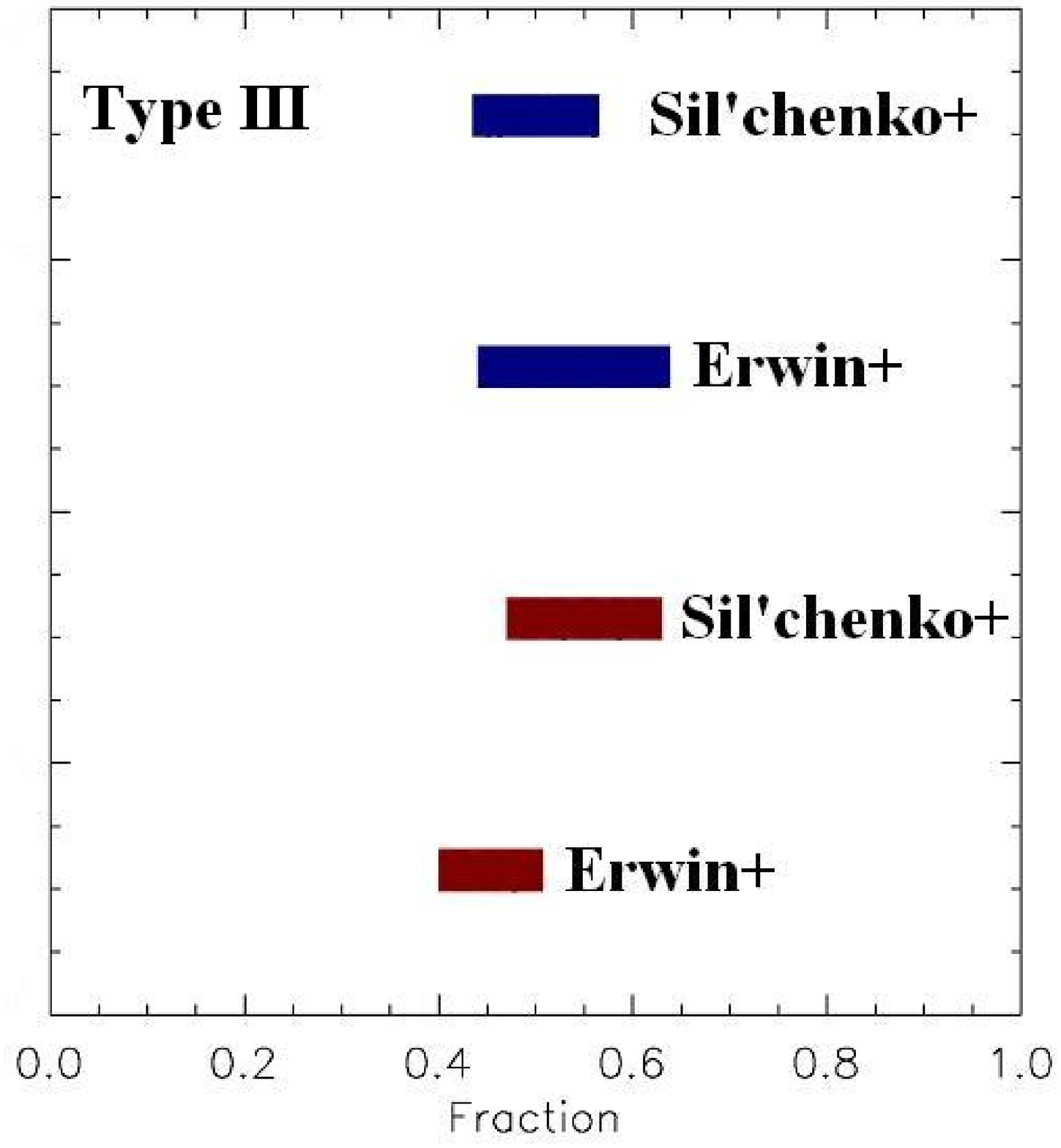} \\
\end{tabular}
\caption{The statistics of the disk surface-brightness profile types in different environments:
Type I in the left, Type II in the middle, and Type III in the right. The position along
the abscissa marks the disk type fraction in a sample considered; the shift along the ordinate
is arbitrary. The width of the bars indicates the estimate accuracy at one sigma level;
the errors shown are derived from the binomial distributions.
The red bars are plotted for the field, and blue bars are plotted for the clusters. The data
are taken from: \citet{erwin12}, marked by 'Erwin+', and \citet{lcogt_clust} and the present
paper, marked by 'Sil'chenko+'.}
\label{envcomp}
\end{figure*}

\citet{erwin12} were the first who reported about the difference between typical structure of
the S0 disks in the field and in a cluster: they had analyzed radial structure of 24 S0 galaxies
-- members of the Virgo cluster -- and had concluded that the statistics of the surface-brightness profile types
in the cluster differed significantly from that in the field. They had not found any Type-II
profiles in the Virgo S0s at all while in the field a quarter of all S0s demonstrated
truncated stellar disks \citep{erwin08,guti_erwin}. The following fractions of the profile types
in the Virgo were reported by \citet{erwin12}: 46\%$\pm 10$\%\ of the Type I, 0\%$\pm 4$\%\ of the Type II,
and the remaining 54\%\ profiles of the Type III. Our results \citep{lcogt_clust} on 60 lenticular galaxies
in 8 southern clusters were: 27 S0s of Type I -- 45\%$\pm 6$\%, 3 S0s of Type II -- 5\%$\pm 3$\%\ (the errors indicated
the root square of the binomial distribution variance), the rest half of all S0s in clusters -- of Type III.
We pointed out then that our results were completely consistent with the statistics of the Virgo S0s
reported by \citet{erwin12} and also differed from the field statistics where 26\%$\pm 6$\%\ of
Type I and 28\%$\pm 6$\%\ of Type II were reported by \citet{erwin12}. Our present results for the quite
isolated S0s are: 12 S0s of Type I -- 28\%$\pm 7$\%, 7 S0s of Type II -- 17\%$\pm 6$\%,
23 S0s of Type III -- 55\%$\pm 8$\%\ (the fraction errors are again estimated as the root square of the binomial
distribution variance). We conclude that the profile type distributions among the isolated S0s does not differ
from that for the field, while mostly group S0s were reported by \citet{erwin08,guti_erwin}. If we compare the present statistics for
the isolated S0s with the results obtained by us with the same methods over the similar data for the cluster S0s
in our previous work \citep{lcogt_clust}, we confirm that the main difference between the cluster S0s and S0s in more
rarefied environments is the depletion of Type II profile occurrence in clusters and its noticeable presence in
sparse environments. The occurrence of Type III profiles in S0 galaxies does not depend on environments -- they
constitute about a half everywhere. All this statistics is visualized in Fig.~\ref{envcomp}.

\subsection{Vertical structure of the S0 disks}

The novel point of our photometric analysis is individual estimates of the stellar disk
relative thicknesses expressed in the terms of \citet{hubble26}'s parameter $q$  -- we present them in 
the Table~\ref{s0type1}, the Table~\ref{s0type2}, and the Table~\ref{s0type3}, for the inner segments of
the piecewise exponential disks.

\begin{figure*}[t]
\centering
\begin{tabular}{c c}
 \includegraphics[width=8cm]{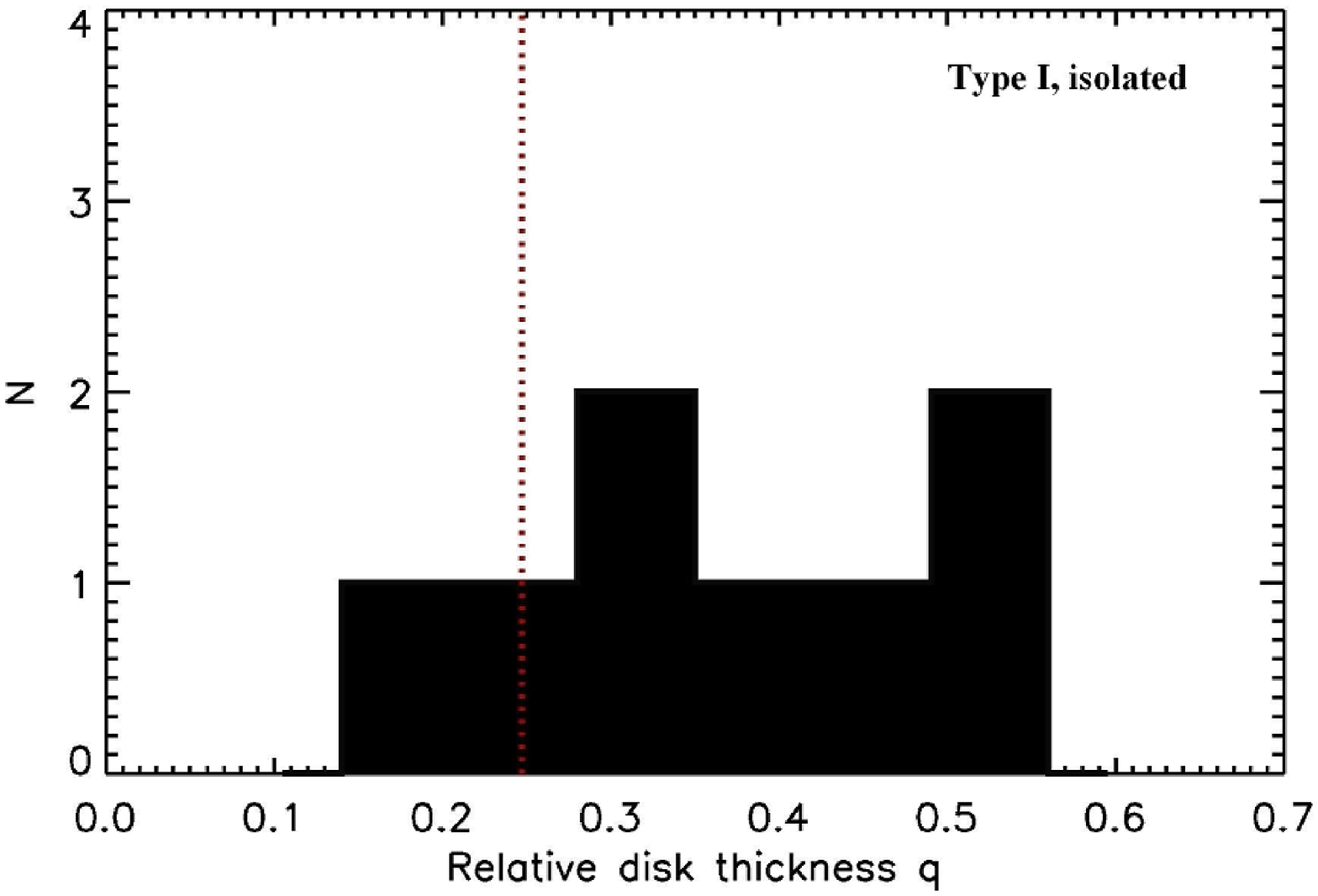} &
 \includegraphics[width=8cm]{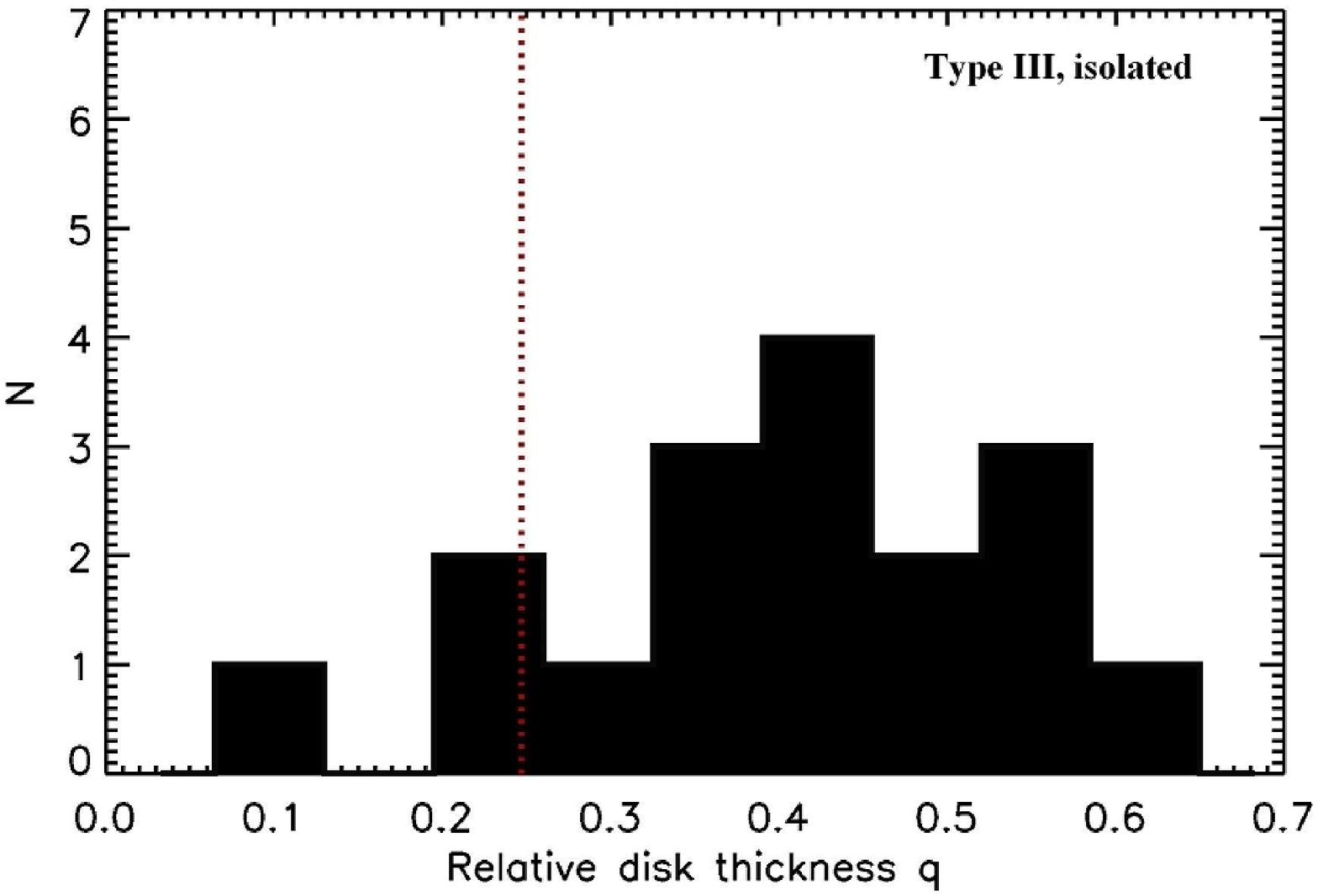} \\
\includegraphics[width=8cm]{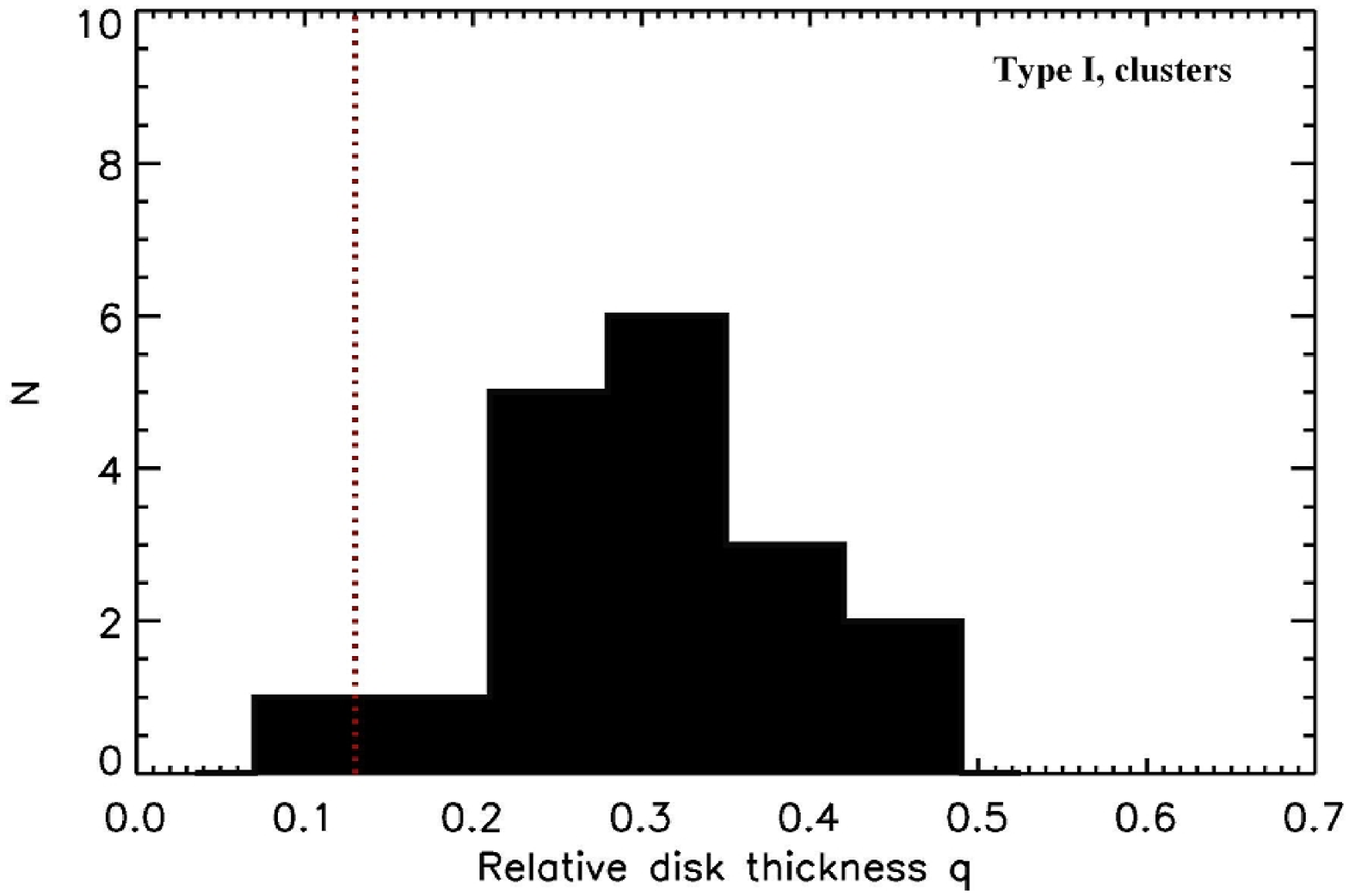} &
 \includegraphics[width=8cm]{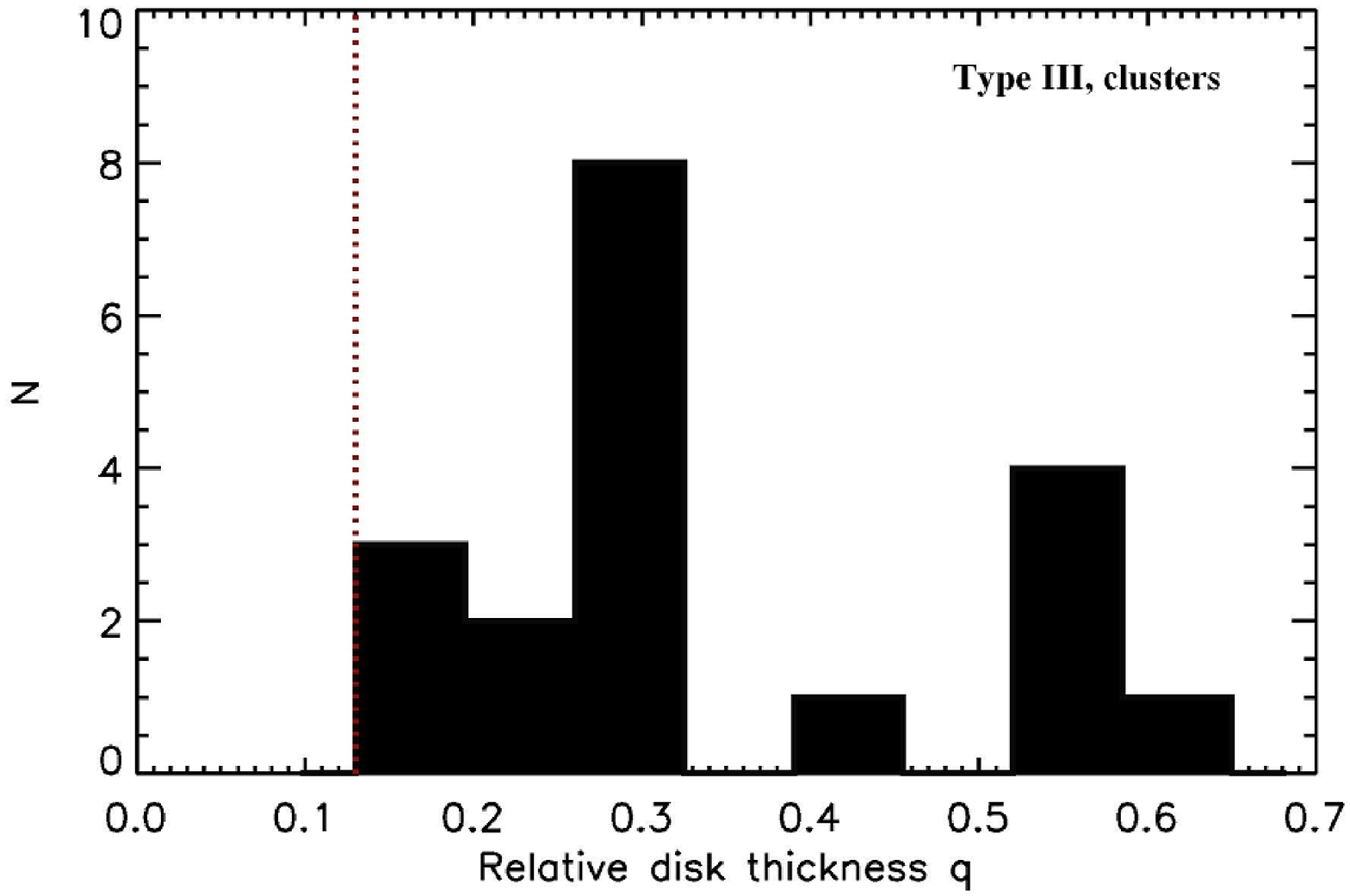} \\
\end{tabular}
\caption{The distributions of the disk relative thickness for the Type-I disks ({\it left}) and
for the inner segments of the Type-III disks ({\it right}). The vertical dotted red lines indicate the mean
thicknesses of the inner segments of the Type-II disks. The upper row shows the results for the isolated S0s,
and the bottom row -- for the cluster S0s from our previous paper \citep{lcogt_clust}.}
\label{qhisto}
\end{figure*}

Figure~\ref{qhisto} presents the distributions of the relative disk thicknesses for the Type-I disks ({\it left})
and for the inner segments of the Type-III disks ({\it right}); in this Figure we compare the thickness distributions
for the isolated S0s (top plots) and for the cluster S0s (bottom plots), the latter obtained in our previous work \citep{lcogt_clust}.
The median thickness of the isolated Type-I disks is 0.35, while the median thickness of the cluster
Type-I disks is 0.31. The cause of this formal dissimilarity is the fact that
the thickness distribution of the isolated S0s Type-I disks includes a few very thick, $q\approx 0.5$, disks
which are absent in clusters. However, the Type-I sample is small hence we cannot made any certain conclusions
about its thickness distribution.
More certain difference is demonstrated by the inner Type-III disks: without pseudobulges, defined as having $q>0.5$, the mean
thickness of the cluster S0 disks is $0.25\pm 0.02$, while the isolated S0 disks have $0.35\pm 0.03$.
The sample of the inner Type-III disks with the measured thicknesses is large enough, so we can apply the K-S test.
By considering the full $q$ distribution for the isolated inner Type-III disks in comparison with the analogous
distribution for the cluster S0s, we find that the K-S statistics is $\lambda=1.28$. It means that the
distributions are different at the 92\%\ confidence level.
The thickness distribution for the inner segments of Type-III disks in the isolated S0s reveals an additional strong peak 
at $q\approx 0.42$ which is absent in the thickness distribution of the cluster Type-III disks. Meanwhile the thin disks
similar to the inner segments of the Type-III disks in clusters are also present in some isolated S0s.
We conclude that additional dynamical mechanisms affecting isolated S0s but not working in clusters may broaden
the vertical structure of the stellar disks in our sample S0s.

\begin{figure}
\centering
\includegraphics[width=0.45\textwidth]{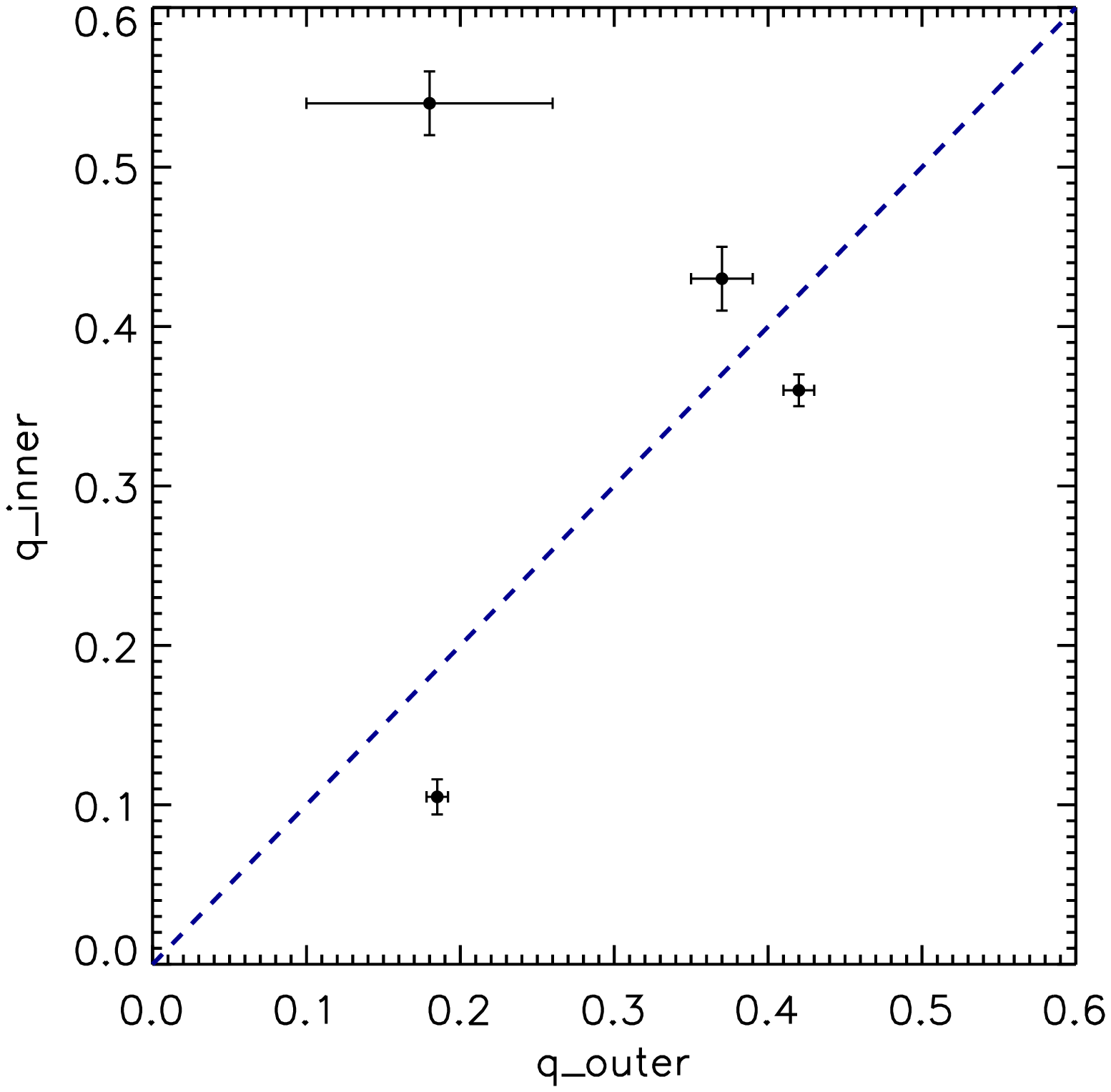}
\caption{The comparison of the thicknesses of the inner and outer exponential segments for the Type-III disks
of the isolated S0s.}
\label{q_corr}
\end{figure}

Interestingly, in a few cases when we are able to estimate the thickness both for the inner and outer segments of
the Type-III disks, they are similar: either both the inner and outer segments are thin, or both are thick (Fig.~\ref{q_corr}).
Among four Type-III disks with the measured thicknesses of both segments, only NGC~1656 has the inner segment much thicker
than the outer one, $q_{inn}=0.54\pm 0.02$ vs $q_{out}=0.18\pm 0.08$; evidently, it is the case of a pseudobulge with the exponential
surface-brightness profile mimicking the anti-truncated disk. But otherwise the thickness of the Type-III disks in the isolated
S0s keeps nearly constant beyond the breaks.

\subsection{Discussion: Particular mechanisms of S0 galaxy evolution in sparse environments}

It is a well-established observational paradox: early-type galaxies in rarefied environments demonstrate signatures
of interaction< minor merging, and/or gas accretion much more often than galaxies in clusters. The most prominent manifestation
of this phenomenon is gas kinematics. For example, the volume-complete
survey of nearby early-type galaxies ATLAS-3D revealed frequent misalignment of rotation planes between gas and stars just in 
S0 galaxies in loose groups and in the field while the Virgo cluster S0s, if they possess gas, showed always coupled rotation 
of stars and gas \citep{atlas3d_10}. We have fulfiled spectral study of the kinematics and stellar populations in the disks of quite
isolated S0s selected from the same parent sample as the present study targets and have found ionized-gas presence
in 75\%\ of galaxies; the gas counterrotates the stars along the spectrograph slit
in the half of the objects analyzed \citep{isomnras}. In addition, the disk stellar populations in the isolated S0s are
in average younger than those of the S0 disks in groups and in clusters \citep{isosalt}.

Now we can look at the structure of the isolated S0s in comparison with the structure of the cluster S0s studied by us in
the previous work \citep{lcogt_clust}. If we compare the bar
frequency by using the data of Table~\ref{listgal} with the analogous Table from the paper by \citet{lcogt_clust}, we can 
find that in the isolated S0s the bars are more frequent than in the cluster S0s:  43\%$\pm 8$\%\ against 27\%$\pm 6$\%.
We may then suggest that the excess of bars in isolated S0s is caused by interactions. All thick inner segments of the Type-III disks 
in our isolated S0s, $q>0.43$, are found in the galaxies with the bars or with signatures of recent minor merging
(shells, blue compact inclusions). This fact can explain why we do not see such thick inner disks in the Type-III cluster S0s.
Minor merging, especially dry minor merging, is a recognized
dynamical mechanism to make stellar disks thick \citep{walker96} and hot \citep{tapia14}; in addition, simulations demonstrate
that just dry minor merging can produce antitruncated radial brightness distributions in the disks \citep{younger}. Otherwise,
the very presence of a strong bar can also produce an antitruncated density profile of a disk through enhanced radial
migration \citep{herpich17}. These are possible extra-mechanisms that can work for the isolated galaxies and can thicken their
stellar disks.

However, the question remains to be open whether there is an evolutionary link between different types of the
disk surface-brightness profiles. While the large-scale cosmological simulations imply transformations, say,
of Type II into Type I and then perhaps into Type III \citep{ruizlara}, the observations of Type-III galaxies
over a range of redshifts reveal quite stable structure parameters and scaling relations just
for this disk type \citep{borlaff18}. Perhaps there is no evolutionary link between disk profile types, and
we must search for initial conditions determining the disk type. Then the environment density
is the most promising candidate for such conditions.

\section{Conclusions}

By undertaking two-band photometric observations at the Las Cumbres Observatory 1m-telescope network,
we have studied large-scale disk structures for the sample of 42 isolated lenticular galaxies of the southern sky.
In our sample we have identified all three types of the radial surface-brightness profiles -- single-scale pure
exponential (Type I), truncated (Type II), and antitruncated (Type III). The last are the most numerous:
about a half of all lenticular galaxies have antitruncated radial brightness profiles, both in isolation and in clusters.
The difference of the S0 disk radial structures between the galaxies evolving in visible isolation and those in clusters consists
of a prominent presence of truncated stellar disks in the isolated S0s -- slightly less than 20\%\ isolated S0s demonstrate
downward brightness breaks in their disks while in clusters such type of radial brightness profiles is almost absent.
We have also measured the individual stellar-disk relative thicknesses for a significant part of our sample by exploring our
original method applicable to the exponential, or partly exponential, disks with arbitrary orientation in the space.
The thickness distributions seems to be different for the Type-III isolated S0s and S0s in clusters: in average,
inner stellar disks of the Type-III isolated S0s are thicker. By summarizing the differences in the vertical and
radial structures of the disks of lenticular galaxies in clusters and in isolation, we conclude that isolated S0s experience dynamical
and structure evolution provoked by some mechanisms absent in clusters.

\acknowledgments

This work is based on the imaging data obtained with the LCO robotic telescope network.
OKS and EMCh acknowledge the support from the Program of development of M.V.
Lomonosov Moscow State University (Leading Scientific School 'Physics of
stars, relativistic objects and galaxies').
AYK acknowledges support from the National Research Foundation of South Africa (NRF).
We have made actively the usage of the HyperLEDA database (http://leda.univ-lyon1.fr). This research
has made use of the NASA/IPAC Extragalactic Database (NED) which is operated by the Jet
Propulsion Laboratory, California Institute of Technology, under contract with the National Aeronautics
and Space Administration. For the purpose of our photometric calibration we have used SDSS/DR9 \citep{sdss_dr9} and 
Pan-STARRS1 data \citep{panstarrs_1,panstarrs_2}.
Funding for the SDSS-III has been provided by the Alfred P. Sloan Foundation, the Participating Institutions,
the National Science Foundation, and the U.S. Department of Energy Office of Science. The SDSS-III Web site is http://www.sdss3.org/.
SDSS-III is managed by the Astrophysical Research Consortium for the Participating Institutions of the SDSS-III Collaboration 
including the University of Arizona, the Brazilian Participation Group, Brookhaven National Laboratory, Carnegie Mellon University,
University of Florida, the French Participation Group, the German Participation Group, Harvard University, the Instituto de 
Astrofisica de Canarias, the Michigan State/Notre Dame/JINA Participation Group, Johns Hopkins University, 
Lawrence Berkeley National Laboratory, Max Planck Institute for Astrophysics, Max Planck Institute for Extraterrestrial Physics, 
New Mexico State University, New York University, Ohio State University, Pennsylvania State University, University of Portsmouth, 
Princeton University, the Spanish Participation Group, University of Tokyo, University of Utah, Vanderbilt University, 
University of Virginia, University of Washington, and Yale University. 
The Pan-STARRS1 Surveys (PS1) and the PS1 public science archive have been made possible through contributions 
by the Institute for Astronomy, the University of Hawaii, the Pan-STARRS Project Office, the Max-Planck Society 
and its participating institutes, the Max Planck Institute for Astronomy, Heidelberg and the Max Planck Institute 
for Extraterrestrial Physics, Garching, The Johns Hopkins University, Durham University, the University of Edinburgh, 
the Queen's University Belfast, the Harvard-Smithsonian Center for Astrophysics, the Las Cumbres Observatory Global 
Telescope Network Incorporated, the National Central University of Taiwan, the Space Telescope Science Institute, 
the National Aeronautics and Space Administration under Grant No. NNX08AR22G issued through the Planetary Science 
Division of the NASA Science Mission Directorate, the National Science Foundation Grant No. AST-1238877, the University 
of Maryland, Eotvos Lorand University (ELTE), the Los Alamos National Laboratory, and the Gordon and Betty Moore Foundation.

\vspace{5mm}
\facilities{LCO, SDSS, Pan-STARRS, NED, HyperLEDA}

\appendix

\section{Observational details}

\startlongtable
\begin{deluxetable}{llrccc}
\tablecolumns{6}
\tablewidth{0pc}
\tablecaption{The galaxies studied photometrically with the LCO network.\label{obslog}}
\tablehead{
\colhead{Galaxy} & \colhead{Telescope}  & \colhead{Date} &
\colhead{Band} & \colhead{Exposure, sec} & \colhead{Seeing,$^{\prime \prime}$} }
\startdata
ESO 003-G001 & Siding Spring, 1m0-03  & 20170901 & g & 800x3 & 2.8 \\
ESO 003-G001 & Siding Spring, 1m0-03  & 20170901 & r & 600x3 & 2.5 \\
ESO 040-G002 & Cerro Tololo, 1m0-05 & 20180220 & g & 800x3 & 1.8 \\
ESO 052-G014 & Siding Spring, 1m0-11 & 20170901 & g & 800x3 & 1.9 \\
ESO 052-G014 & Siding Spring, 1m0-11  & 20170901 & r & 600x3 & 1.8 \\
ESO 069-G001 &  Siding Spring, 1m0-03 & 20170901 & g & 800x3 & 2.4 \\
ESO 069-G001 &  Siding Spring, 1m0-03 & 20170901 & r & 600x3 & 2.4 \\
ESO 235-G051 & SAAO 1m0-12 & 20170926 & g & 800x3 & 1.3 \\
ESO 235-G051 & SAAO 1m0-12 & 20170926 & r & 600x3 & 1.1 \\
ESO 265-G033 & Cerro Tololo, 1m0-04 & 20160410 & g & 800x3 & 1.8 \\
ESO 265-G033 & Cerro Tololo, 1m0-04 & 20160410 & r & 600x3 & 1.3\\
ESO 269-G013 & Cerro Tololo, 1m0-09 & 20160408 & g  & 800x3 & 1.7\\
ESO 269-G013 & Cerro Tololo, 1m0-09 & 20160408 & r  & 600x3 & 1.4\\
ESO 274-G017 & Siding Spring, 1m0-03 & 20170920 & g & 800x3 & 1.8 \\
ESO 274-G017 & Siding Spring, 1m0-03 & 20170920 & r & 600x3 & 1.8 \\
ESO 316-G013 & SAAO 1m0-10 & 20171224 & g & 800x3 & 1.6 \\
ESO 316-G013 & SAAO 1m0-10 & 20171224 & r & 600x3 &  1.9 \\
ESO 324-G029 & Cerro Tololo, 1m0-04 & 20160408 & g  & 800x3 & 1.8 \\
ESO 324-G029 & Cerro Tololo, 1m0-04 & 20160408 & r  & 600x3 & 1.9 \\
ESO 446-G049 & Cerro Tololo, 1m0-04  & 20180120 & g & 800x3 & 1.9 \\
ESO 446-G049 & Cerro Tololo, 1m0-04  & 20180120 & r & 600x3 & 1.7 \\
ESO 469-G006 & Cerro Tololo, 1m0-04  & 20160807  & g & 800x3 &  1.9 \\
ESO 469-G006 & Cerro Tololo, 1m0-04  & 20160807  & r & 600x3 &  1.8 \\
ESO 486-G038 & Siding Spring, 1m0-03 & 20170901 & g & 800x3 & 2.2 \\
ESO 486-G038 & Siding Spring, 1m0-03 & 20170901 & r & 600x3 & 2.2 \\
ESO 496-G003 & Cerro Tololo, 1m0-05  & 20171223  &  g  & 800x3  & 1.5 \\
ESO 496-G003 & Cerro Tololo, 1m0-05  & 20171223  &  r  & 600x3  & 1.4 \\
ESO 506-G011 & Cerro Tololo, 1m0-09  & 20160409 &  g  & 800x3  & 1.6 \\
ESO 506-G011 & Cerro Tololo, 1m0-09  & 20160409 &  r  & 600x3  &  1.5 \\
ESO 508-G033 & Cerro Tololo, 1m0-09  & 20160408  &  g  & 800x3  & 1.5  \\
ESO 508-G033 & Cerro Tololo, 1m0-09  & 20160408  &  r  & 600x3  & 1.8  \\
ESO 545-G040 & Siding Spring, 1m0-03 & 20170901 & g & 800x2 & 2.1 \\
ESO 545-G040 & Siding Spring, 1m0-03 & 20170901 & r & 600x3 & 2.05 \\
ESO 563-G024 &  SAAO 1m0-10 & 20171215  &  g  & 800x3  & 1.7 \\
ESO 563-G024 &  SAAO 1m0-10 & 20171215  &  r  & 600x3  & 1.4 \\
ESO 603-G029 &  Cerro Tololo, 1m0-05  & 20171215  &  g  & 800x3  & 1.9  \\
ESO 603-G029 &  Cerro Tololo, 1m0-05  & 20171215  &  r  & 600x3  & 1.6 \\
IC 276 &  Siding Spring, 1m0-11 & 20170901 & g & 800x2 & 2.1 \\
IC 276 &  Siding Spring, 1m0-11 & 20170901 & r & 600x2 & 1.9 \\
IC 537 & SAAO, 1m0-12  & 20180221   &  g  & 800x3   & 2.4 \\
IC 537 & SAAO, 1m0-12  & 20180221   &  r  & 600x2   & 2.3 \\
IC 537 & Cerro Tololo, 1m0-04  &  20180321  &  g & 800x3 & 2.3 \\
IC 4913  & Cerro Tololo, 1m0-04  & 20160410 & g & 800x3 & 1.5 \\
IC 4913 & Cerro Tololo, 1m0-04  & 20160410 & r & 600x3 & 1.4 \\
NGC 270 & McDonald Observatory, 1m0-08 & 20170831 & g & 800x3 & 1.7 \\
NGC 270 & McDonald Observatory, 1m0-08 & 20170831 & r & 600x3 & 1.5 \\
NGC 324 &  Cerro Tololo, 1m0-04 & 20170923 & g & 800x3 & 1.7 \\
NGC 324 &  Cerro Tololo, 1m0-04 & 20170923 & r & 600x3 & 1.5 \\
NGC 1656 & Siding Spring, 1m0-11 & 20170925 & g &  800x3 & 3.1 \\
NGC 1656 & Siding Spring, 1m0-11 & 20170925 & r &  600x3 & 2.8 \\
NGC 4087  & Cerro Tololo, 1m0-05  & 20171223 & g & 800x3 & 1.6 \\
NGC 4087  & Cerro Tololo, 1m0-05  & 20171223 & r & 600x3 & 1.3 \\
NGC 4878 &  McDonald Observatory, 1m0-08 & 20180220 & g & 800x2 & 2.7 \\
NGC 4878  & McDonald Observatory, 1m0-08 & 20180321 & g & 800x3 & 1.7 \\
NGC 4878  & McDonald Observatory, 1m0-08 & 20180321 & r & 600x3 & 1.5 \\
NGC 5890  &  McDonald Observatory, 1m0-08  & 20180321 & g & 800x3 & 2.1 \\
NGC 5890  &  McDonald Observatory, 1m0-08  & 20180321 & r & 600x3 & 1.6 \\
NGC 6014  & Cerro Tololo, 1m0-05  & 20180321 & g & 800x3 & 1.9 \\
NGC 6014  & Cerro Tololo, 1m0-05  & 20180321 & r & 600x3 & 1.4 \\
NGC 7007  & Siding Spring, 1m0-03  & 20170901 & g & 800x3 & 2.4 \\
NGC 7007  & Siding Spring, 1m0-03  & 20170901 & r & 600x3 & 2.2 \\
NGC 7208  & Cerro Tololo, 1m0-09  & 20160808 & g & 800x3 & 1.9  \\
NGC 7208  & Cerro Tololo, 1m0-09  & 20160808 & r & 600x3 & 2.0 \\
PGC 11756  &  McDonald Observatory, 1m0-08 & 20170831 & g & 800x3 & 1.6 \\
PGC 11756  &  McDonald Observatory, 1m0-08 & 20170831 & r & 600x3 & 1.5 \\
PGC 16688  &  Cerro Tololo, 1m0-05 & 20170925 & g & 800x3 & 1.6 \\
PGC 16688  &  Cerro Tololo, 1m0-05 & 20170925 & r & 600x3 & 1.6 \\
PGC 34728  &  Siding Spring, 1m0-03  & 20171223  & r & 600x3  & 1.9 \\
PGC 35771  & Cerro Tololo, 1m0-09   & 20160410   &   g  & 800x3  & 2.5  \\
PGC 35771  & Cerro Tololo, 1m0-09   & 20160410   &   r  & 600x3  & 2.0  \\
PGC 46474  & Cerro Tololo, 1m0-04   & 20180319  &   g  & 800x3  & 1.6 \\
PGC 46474  & Cerro Tololo, 1m0-04  &  20180319  &   r  & 600x3  & 1.5 \\
PGC 52002  &  Cerro Tololo, 1m0-05 & 20180321  &  g  & 800x3  &  1.4 \\
PGC 52002  &  Cerro Tololo, 1m0-05 & 20180321  &  r  & 600x3  & 1.4 \\
PGC 58114  &  McDonald Observatory, 1m0-08  & 20180321 & g & 800x3  & 1.5 \\
PGC 58114  &  McDonald Observatory, 1m0-08  & 20180321 & r & 600x3  & 1.5 \\
PGC 63536  & SAAO 1m0-12  & 20180320 & g & 800x3 & 2.1 \\
PGC 63536  & SAAO 1m0-12  & 20180320 & r & 600x3 & 2.1 \\
PGC 68401  & Cerro Tololo, 1m0-04  & 20160808 &  g  & 800x3  & 1.9 \\
PGC 68401  & Cerro Tololo, 1m0-04  & 20160808 &  r  & 600x3  &  1.8 \\
UGC 3097   & SAAO 1m0-10  & 20171210 & g & 800x2 & 2.5 \\
UGC 3097   &  SAAO 1m0-10 & 20171224 & g & 800x3 & 1.8 \\
UGC 3097   &  SAAO 1m0-10 & 20171224 & r & 600x3 & 1.5 \\
UGC 5745   & Cerro Tololo, 1m0-04  & 20180220 & g & 800x3 & 1.9 \\
UGC 5745   & Cerro Tololo, 1m0-04  & 20180220 & r & 600x3 & 1.7 \\
\enddata
\end{deluxetable}

\end{document}